\documentstyle[prl,aps,twocolumn,epsf]{revtex}

\begin{document}
\twocolumn[\hsize\textwidth\columnwidth\hsize\csname @twocolumnfalse\endcsname
\title{Dirac Nodes and Quantized Thermal Hall
Effect in the Mixed State of d-wave Superconductors}
\author{Ashvin Vishwanath}
\address{Department of Physics, Joseph Henry Laboratories, Princeton University, Princeton, NJ 08544.}
\date{\today}
\maketitle 
\begin{abstract}
We consider the vortex state of d-wave superconductors in the clean limit. Within the linearized approximation the quasiparticle bands obtained are found to posess Dirac cone dispersion (band touchings) at special points in the Brillouin zone. They are protected by a symmetry of the linearized Hamiltonian that we call ${\mathcal T}_{Dirac}$. Moreover, for vortex lattices  that posess inversion symmetry, it is shown that there is always a Dirac cone centered at zero energy within the linearized theory. On going beyond the linearized approximation and including the effect of the smaller curvature terms (that break ${\mathcal T}_{Dirac}$), the Dirac cone dispersions are found to acquire small gaps  ($\approx 0.5$KTesla$^{-1}$ in YBCO) that scale linearly with the applied magnetic field. When the chemical potential for quasiparticles lies within the gap, quantization of the thermal-Hall conductivity is expected at low temperatures i.e. $\frac{\kappa_{xy}}{T} = n(\frac{\pi^2k_B^2}{3h})$ with the integer $n$ taking on values $n=\pm 2,0$. This quantization could be seen in low temperature thermal transport measurements of clean d-wave superconductors with good vortex lattices.
\end{abstract}
\vspace{0.15in}
]
\newcommand{\fig}[2]{\includegraphics[width=#1]{#2}}
\newcommand{\dxsq}{d_{x^2-y^2}}
\newcommand{\be}{\begin{equation}}
\newcommand{\ee}{\end{equation}}
\renewcommand{\slash}[1]{\mbox{$\not\!#1\/$}}
\newcommand{\kslash}{\slash{k}}
\newcommand{\sigmaz}{\mbox{\boldmath $\sigma_z$}}
\newcommand{\sigmax}{\mbox{\boldmath $\sigma_x$}}
\newcommand{\sigmaplus}{\mbox{\boldmath $\sigma^+$}}
\newcommand{\TAU}{\mbox{\boldmath $\tau$}}
\newcommand{\alpher}{\alpha^{-1}}
\newcommand{\tpsi}{\tilde{\psi}}
\newcommand{\tV}{\tilde{V}}
\newcommand{\ta}{\tilde{a}}
\newcommand{\One}{\mbox{\bf 1}}
\newcommand{\done}{d_1(\vec{r})}
\newcommand{\dtwo}{d_2(\vec{r})}
\newcommand{\Psx}{P_{sx}(\vec{r})}
\newcommand{\Ps}{\vec{P}_{s}}
\newcommand{\hcbye}{$\frac{hc}{e}$}
\newcommand{\hcbytwoe}{$\frac{hc}{2e}$}
\newcommand{\r}{\vec{r}}
\newcommand{\Tdirac}{$\mathcal{T}_{\mathrm{Dirac}}$}
\newcommand{\sgn}[1]{\mbox{sgn({$#1$})}}

\section{Introduction}
Since the experimental verification of the d$_{x^2-y^2}$ nature of superconductivity in the cuprate materials \cite{Harlingen} there has been much activity in studying the physics of quasiparticles in a d-wave superconductor. In contrast to the s-wave case, the d-wave superconducting gap vanishes at points on the Fermi surface leading to low energy quasiparticles that behave like massless Dirac fermions. One question that is of great interest is the behaviour of these quasiparticles in the mixed state of superconductors - a problem that is both theoretically rich as well as of relevance to explaining several different experiments. When the vortices are arranged in a lattice, the situation is reminescent of the Hofstadter problem of charged particles in a magnetic field, subject to a periodic potential, where much interesting physics is known to emerge. Here, however, the periodic lattice is {\it generated} by the magnetic field, and, as a result of flux quantization, is always in commensuration with the field. The quasiparticles, which do not carry a definite charge, also couple differently to the field - in fact they only couple to the superflow arising from the combined effect of the magnetic field as well as the vortices. Furthermore, there is a statistical interaction between $hc/2e$ vortices and quasiparticles, which causes them to change sign on circling a vortex. As we shall see, these properties will introduce interesting new features into the problem. Experimentally, there have been several probes of quasiparticle behaviour in a magnetic field, including measurement of thermal hall conductivity $\kappa_{xy}$ \cite{Ong0,Ong} and low temperature longitudinal thermal conductivity $\kappa_{xx}$ \cite{Taillefer0,Taillefer} and specific heat \cite{Junod}. A satisfactory explanation of the results of these experiments in terms of d$_{x^2-y^2}$ quasiparticles would be strong support for the point of view that the superconducting phase of the cuprates is a conventional d-wave superconductor.  On the other hand, if such an explanation proves elusive, it may well point to some additional, and perhaps exotic, physics even in the superconducting state of the cuprates. 

	This problem of Dirac quasiparticles in the mixed state has been considered by several authors. Gorkov and Schriffer \cite{Gorkov&Scriffer}, and Anderson \cite{Anderson} proposed a Landau level like spectrum. In the latter work, this was derived by mapping the problem onto a Dirac particle in a uniform magnetic field, under the assumption that the superflow may be neglected. In \cite{Morita}, Morita et al. studied a lattice model of the problem, first in the same approximation as \cite{Anderson}, and then numerically with the superflow included. Topological aspects of relevance to this problem were pointed out. 

	A different approach was taken by Franz and Tesanovic who considered the problem within the linearized approximation and introduced a ficticious U(1) gauge field to implement the statistical interaction between the vortices and quasiparticles (Franz-Tesanovic transformation) \cite{FT}. This led to a numerical evaluation of the resulting quasiparticle band structure, which was extended by the detailed study of Marinelli et al. \cite{Marinelli}, Vafek et al. \cite{Vafek} and Kallin et al. \cite{Kallin}. 

	Here too we begin by considering the the problem of d-wave quasiparticles in the vortex lattice state within the linearized approximation. In contrast to some earlier approaches, we focus on identifying the symmetries of the problem and their consequences for the spectrum of the linearized Hamiltonian. Our results are easily stated - the Hamiltonian, as a consequence of linearization, posesses an additional symmetry (which we call ${\mathcal T}_{Dirac}$) that preserves Dirac cones at certain special points in the Brilloin zone. The energy dispersion, in the vicinity of these points, is that of a massless Dirac particle. If the vortex lattice posesses inversion symmetry, then there is a Dirac cone centered at {\it zero energy}. 

While these results are in general agreement with the results obtained in earlier numerical work \cite{FT,Marinelli,Vafek,Kallin} we point out out a subtle feature of the Franz-Tesanovic transformation in the linearized approximation, which could yield spurious gaps in numerical simulations and which we believe to be the cause for discrepancies from the results derived here. 

	We then proceed to consider the effect of the curvature terms, that were dropped when linearizing the Hamiltonian. These terms arise, for example, from the parabolic nature of the electron dispersion. In the parameter range of interest they may be considered as small perturbations on the linearized Hamiltonian. However, as pointed out in \cite{Simon&Lee}, they are crucial to generating a non-vanishing thermal Hall response. Since the curvature terms break the symmetry    ${\mathcal T}_{Dirac}$ that protects the Dirac nodes, they give rise to a small gap at these nodes, and the energy dispersion in their vicinty is now that of a massive Dirac particle. Thus for temperatures that are  smaller than this gap scale, the curvature terms can have a qualitative effect on the properties of the system. 


	In addition to information about the quasiparticle spectrum, it is in some cases possible within our approach to derive consequences for low temperature thermal transport. To this end, we first recall that massive Dirac particles in two dimensions exhibit a quantized Hall effect when the chemical potential lies in the gap \cite{Haldane},\cite{mpaf}. In the presence of a vortex lattice with inversion symmetry, the chemical potential for the superconductor quasiparticles will lie at the center of the gap induced by the curvature terms, if we ignore Zeeman splitting. As a result, a quantized thermal Hall conductance of $\kappa_{xy}/T = \pm 1/2$ (in appropriate units, at low temperatures) is expected from each of the four nodes, and can give rise to two scenarios $\kappa_{xy}/T = \pm 2$ or $0$, while $\kappa_{xx}/T \rightarrow 0$ in both cases. First, if the contribution of all four nodes is of the same sign, $\kappa_{xy}/T = \pm 2$, we have a nontrivial quantized thermal Hall conductance that is expected to be attained for temperatures smaller than the gap. This situation is topologically identical to (i.e. has the same edge content as) a {\it pure} d$_{x^2-y^2}$ + i d$_{xy}$ superconductor, that is also known to exhibit a quantized  thermal hall effect \cite{Volovik}\cite{Senthil}. A magnetic induction of such a pairing symmetry in the cuprates was proposed in \cite{Laughlin}\cite{TVR}; here we will provide a concrete realization of these general ideas, and layout the route to calculating physical parameters such as, for example, the size of the energy gap. The second scenario is when the Hall conductances from the different nodes cancel leading to a $\kappa_{xy}/T = 0$, that is topologically identical to a d$_{x^2-y^2}$+is superconductor, or any other thermal insulator. Which of these scenarios is realized is a function of vortex lattice geometry and the anisotropy of the Dirac dispersion in the homogenous superconductor.  In order for the quantization to be visible experimentally the energy gap needs to be larger than the Zeeman splitting, which plays the role of chemical potential for the quasiparticles . Since the energy gap is also found to scale linearly with magnetic field, and is roughly of the same magnitude as the Zeeman energy, the question of which one is larger in any particular material is a detailed quantitative issue.  Here we will demonstrate how fairly simple numerical calculations within our theory can predict in a given situation, which of the scenarios described is realized, as well as provide a quantitative estimate of the energy gap and its dependence on various physical parameters.

	The rest of this paper is organized as follows. In Section II, we begin by laying out our assumptions and then deriving the Bogoliubov-de Gennes equation that describes d-wave quasiparticles in the mixed state. The quasiparticles are shown to couple to the superflow generated by the combined effect of the vortices and the magnetic field. In addition, they acquire a Berry phase of (-1) on circling a $hc/2e$ vortex. We then derive the linearized approximation that describes the low energy quasiparticle excitations in the magnetic field range of interest. These linearized equations describing d-wave quasiparticles in a vortex lattice are analyzed in Section III. For purposes of clarity, we first consider a vortex lattice of $hc/e$ (double) vortices. The simplifying feature here is that the Berry phase terms are absent and we only have to contend with the effects of the superflow. A symmetry of the linearized Hamiltonian that protects the Dirac nodes is identified, and the role of inversion symmetry in maintaining a Dirac node at {\it zero energy} is described. We then tackle the physically more interesting case of a vortex lattice of $hc/2e$ vortices. Here, a similar though more involved analysis obtains for us the same results as for a vortex lattice of $hc/e$ vortices. A check on these arguments within perturbation theory is detailed in Appendix A. In Appendix B, we turn to some subtle aspects of the Franz-Tesanovic (FT) transformation for the linearized problem, especially the issue of gauge invariance under different FT transformations, and the need to properly regularize the linearized theory in order to obtain a faithful representation of the problem. An alternate argument for the existence of Dirac nodes in the linearized problem for the $hc/2e$ vortex lattice case, that does not utilize the Franz-Tesanovic transformation, is also presented in this appendix. A detailed comparison of our results for the linearized theory against earlier numerical work is presented in Appendix C. In Section IV, we go beyond the linearized approximation by including the effect of curvature terms and discuss implications for low temperature heat transport. A numerical calculation of the gaps induced by the curvature terms for the $hc/e$ square vortex lattice case is also presented. Finally, in Section V we conclude with some brief comments on the effect of vortex disorder and a comparison of our theoretical expectations against available experimental data. 

\section{The BdG equations for d-wave Quasiparticles in the Mixed State}
In this section we derive the equations governing quasiparticles in a d$_{x^2-y^2}$ superconductor, in the presence of a vortex lattice. We begin by detailing the assumptions and approximations that we will make in what follows. 

\subsection{Assumptions and Approximations}
(a) {\it Existence of quasiparticles:} The systems that we will mainly be interested in are the cuprate high temperature superconductors, that are known to have d$_{x^2-y^2}$ gap symmetry. We assume that this superconducting state is otherwise conventional, in particular that there are well defined quasiparticle excitations in this phase, for which there is experimental support from angle resolved photoemission studies 
. For inhomogenous situations, such as the vortex lattice state, we assume that the quasiparticles are governed by an appropriate Bogoliubov-de Gennes type equation. 

(b) {\it Neglecting Vortex Core Contributions:} The vortex core is the region around the center of the vortex of size $\xi$, the coherence length, where the magnitude of the order parameter is significantly supressed from its bulk value. At fields much smaller than $H_{c2}$, the vortex cores of extreme type II superconductors are significantly smaller than the separation between vortices. This is the situation, for example, in optimally doped YBa$_2$Cu$_3$O$_{6.9}$ (YBCO) over the accessible field range. The vortex cores have a size $\xi \sim$ 15 A while the inter-vortex separation is of order 500 A in a 1 Tesla field. Since the vortex cores  take up so little of the sample area (0.1\% in this example) we neglect the modulation of the order parameter magnitude, while retaining its phase variation. 
   
(c) {\it Perfect Vortex Lattice:} We will make the assumption that the superconductor is clean and the vortices are arranged in a perfect lattice. Of course, in any real situation, vortex disorder is expected to be present. In some cases it may even be so large as to destroy the long range positional order of the lattice, in which case, of course, the perfect lattice approximation will not be a good starting point - for example, in BSCCO in magnetic fields of around one Tesla, neutron scattering studies indicate that an ordered vortex lattice is absent \cite{BSCO}. However in YBCO, Bragg spots from the vortex lattice are seen \cite{YBCO} for which the starting point of a perfect vortex lattice may be more justified. Despite the existence of vortex disorder, in what follows we shall proceed with the assumption of a perfect vortex lattice as it provides us with a theoretically tractable starting point. Besides, there is some evidence from experiments \cite{Taillefer} and theory \cite{Vekhter&Houghton} that the scattering from vortex disorder at low temperatures is rather small, and neglecting its effect may be permissible at a first approximation. Extending the theory to include the effects of weak vortex disorder is left for future investigation, although we briefly return to the topic of vortex disorder in Section V, while considering the stability of our results.

\subsection{Bogoliubov-de Gennes equations for d-wave superconductors in the mixed state}

Consider a two dimensional d-wave superconductor in the pure state, which may be described by the model Hamiltonian
\be
{\mathcal H}_{pure} =\int_p \sum_\sigma c^\dag_{p\sigma}\epsilon(p)c_{p\sigma} + \frac12 (\Delta(p)c^\dag_{p\sigma} \tau_{\sigma \sigma'}c^\dag_{-p\sigma'} + \mbox{h.c.})
\ee

where $c^\dag_{p\sigma}$ is the creation operator for an electron of momentum $p$ and spin projection $\sigma$ and $\tau_{\sigma \sigma'}$ are elements of the unit antisymmetric matrix:

\be
\TAU = \left( 
\matrix{0 & 1 \cr-1 & 0} 
\right)	
\ee
Here we consider a parabolic dispersion for the electrons:
\be
\epsilon(p) = \frac1{2m}p^2 - E_F
\ee
and a d$_{xy}$ symmetry of the gap function (equivalent to d$_{x^2-y^2}$ rotated by 45$^0$) which could be taken to have, near the Fermi momentum, the functional form:
\be
\Delta(p) = \frac{\Delta_0}{p_F^2}p_x p_y
\ee
. We now consider the effect of a magnetic field applied along the z axis, that gives rise to vortices. It is convenient at this stage to define the `d' operators \cite{Senthil} as:

\begin{eqnarray*}
d_{1p} &=& c_{p\uparrow}\\
d_{2p} &=& c^\dag_{-p \downarrow}
\end{eqnarray*}
here the spin projection is taken in the direction of the magnetic field. The Hamiltonian for a layer of superconductor in the mixed state can them be written as:
\be
{\mathcal H}= \int d^2r 
	\left( 
	\matrix{d^\dag_1(r) & d^\dag_2(r)} 
	\right)
	H_{BdG} 
	\left( 
	\matrix{d_1(r)  \cr d_2(r)}
	\right)
	+ {\mathcal H}_{Zeeman}
\ee
where 
\be
\label{bdg}
H_{BdG} =  \left[ 
		\matrix{\epsilon(-i\vec{\nabla}-e\vec{A}(\vec{r})) & \mbox{e}^{i\frac\phi2}\Delta(-i\vec{\nabla})\mbox{e}^{-i\frac\phi2} \cr \mbox{e}^{-i\frac\phi2}\Delta(-i\vec{\nabla})\mbox{e}^{i\frac\phi2} & -\epsilon(i\vec{\nabla}-e\vec{A}(\vec{r}))} 
	\right]
\ee
where $\vec{A}$ is the vector potential $\vec{\nabla}\times \vec{A}(\vec{r})=B(\vec{r})\hat{z}$ and $\phi(\vec{r})$ is the phase of the order parameter; and
\be
{\mathcal H}_{Zeeman} = \frac{g\mu_B}2 \int d^2r B(\vec{r}) [d^\dag_1(\vec{r})\done+d^\dag_2(\vec{r})\dtwo]
\ee
. The Hamiltonian written in terms of the `d' particles contains no anomalous terms which implies that the total number of `d' particles is conserved. This is simply a reflection of the fact that the z component of the spin is conserved in this situation. Thus, the density of `d' particles, $\rho_D(\vec{r})=(d^\dag_1(\vec{r})\done+d^\dag_2(\vec{r})\dtwo)$ is proportional to the spin density of the quasiparticles along the z axis, and so appears in the expression for the Zeeman coupling to the magnetic field. The uniform part of the magnetic field density will therefore behave as a chemical potential for the `d' particles. In what follows, we shall not retain the Zeeman term, but rather comment on its effect at the very end. 

	In equation (\ref{bdg}) the phase variation of the order parameter, induced by the vortices, has been written in the form $\mbox{e}^{i\frac\phi2}\Delta(-i\vec{\nabla})\mbox{e}^{-i\frac\phi2}$. This preserves gauge invariance and is consistent with the formulae in \cite{Ye,Vafek}. The variation of the magnitude of the order parameter has been dropped, as discussed earlier. We shall take as given the magnetic field distribution $B(\vec{r})$ and the phase variation, which must satisfy the equation
\be
\vec{\nabla}\times(\vec{\nabla} \phi) = 2\pi \sum_i \delta^{(2)}(\vec{r}-\vec{R}_i)\hat{z}
\ee
 where $\vec{R}_i$s denote the positions of the vortices and the sum runs over all vortices in the lattice. 

	Due to the conservation of the `d' particles, and the fact that they do not interact with each other at this level, we can write a wave equation which contains all the physics of this many body system. This wave equation is the Bogoliubov-deGennes equation:
\be
H_{BdG}	\left( 
	\matrix{
	u(r) \cr
	v(r)}
	\right) =
E \left( 
	\matrix{
	u(r)\cr
	v(r) }
\right)
\ee 

where $H_{BdG}$ has been defined in equation (\ref{bdg}) and can be written in the compact form:
$$
H_{BdG}= \epsilon(-i\vec{\nabla}-e\vec{A}(\vec{r})\sigmaz)\sigmaz + \{\mbox{e}^{i\frac{\phi(\vec{r})}2}\Delta(-i\vec{\nabla})\mbox{e}^{-i\frac{\phi(\vec{r})}2}\sigmaplus + \mbox{h.c.}\}
$$

The $\mbox{\boldmath{$\sigma$}}$s are the 2$\times$2 Pauli matrices in the usual representation. (In this language, the Zeeman term takes the form $H_{Zeeman}=\frac{g\mu_B}{2}B(\vec{r})\mbox{\boldmath{$1$}}$).

It is convenient to make a gauge transformation to eliminate the phase variation from the order parameter (London gauge) which may be affected by the unitary transformation:
\be
\label{U}
U=\mbox{e}^{-\frac{i}{2}\phi(\vec{r})\sigmaz}
\ee

The transformed Hamiltonian takes the simple form,
\be
H_{BdG}' = U H_{BdG}U^{-1} = \epsilon(-i\vec{\nabla}+\vec{P}_s(\vec{r})\sigmaz)\sigmaz + \Delta(-i\vec{\nabla}) \sigmax
\ee

where $\vec{P}_s=\frac12 \vec{\nabla}\phi-e\vec{A}$ a, gauge invariant quantity, is the mechanical momentum carried by each member of Cooper pair at point $\vec{r}$. We will sometimes refer to this quantity as the superflow, though this terminology is not quite accurate. Given this definition of $\vec{P}_s$, it is easily seen that
\be
\label{curlps}
\vec{\nabla}\times \vec{P}_s = [\{\frac12 h \sum_i \delta(\vec{r}-\vec{R}_i)\} - eB(\vec{r})]\hat{z}
\ee

In the presence of elementary $\frac{hc}{2e}$ vortices, it must be noted that the unitary transformation (\ref{U}) is not single valued, but changes sign on circling an odd number of such vortices, since it depends on the half angle $\phi(r)/2$ that winds by an odd multiple of $\pi$. Thus the transformed quasiparticle wavefunctions, ($u'$  $v'$)$^{T}$ defined by, 
\be
\label{psiprime}
	\left( 
	\matrix{
	u'(r) \cr
	v'(r) }
	\right) 
	= U
		\left( 
	\matrix{
	u(r) \cr
	v(r)}
	\right) 
\ee

are not single valued, but change sign on circling an odd number of $\frac{hc}{2e}$ vortices. Methods for handling this statistical interaction between quasiparticles and vortices, are described in the next subsection.

Thus, writing the Bogoliubov-deGennes equation in this gauge clarifies how the quasiparticles interact with the magnetic field - which is as summarised below.
\begin{itemize}
\item{The quasiparticles couple to the superflow set up by the combined effect of the vortices and the magnetic field, namely to the combination $\vec{P}_s=\frac12 \vec{\nabla} \phi - e\vec{A}$, and not individually to the phase variation or the vector potential.}
\item{Pick up a (-1)  Berry phase factor on circling a $\frac{hc}{2e}$ vortex.}
\item{Also interact with amplitude variations of the order parameter, which we have dropped.}
\end{itemize}

\begin{figure}
\epsfxsize=2.8in
\centerline{\epsffile{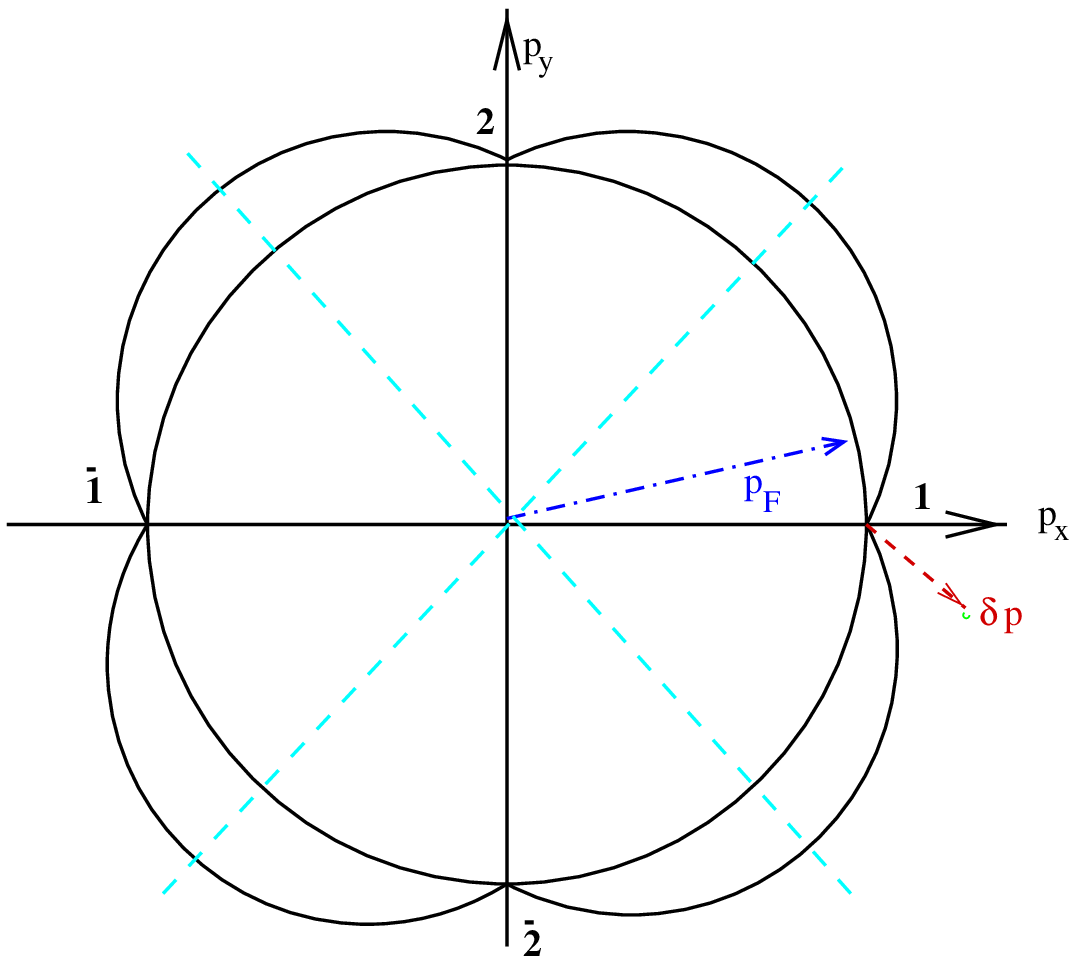}}
\caption{The gap structure for a d$_{xy}$ superconductor with nodes along the coordinate axes.}
\label{nodes}
\end{figure}

\subsection{The Linearized Approximation}

In the limit of low temperatures, and magnetic fields much smaller than H$_{c2}$, we can restrict our attention to the low energy quasiparticle excitations near the gap nodes of the d-wave superconductor. We first recall the linearized approximation for the pure case, which gives us the Dirac dispersion of the nodal quasipartcles. The Hamiltonian for quasiparticles with momentum `p' is $H_p = \epsilon(p) \sigmaz + \Delta(p) \sigmax$. For low energy excitations in the vicinity of node 1 (see Figure \ref{nodes}) this can be expressed in terms of $\vec{p}=p_F\hat{x}+\vec{\delta p}$ (where $\vec{\delta p}$ is measured from node 1), as
\be
H=v_F\delta p_x\sigmaz + v_\Delta \delta p_y \sigmax +\{ \frac1{2m} (\vec{\delta p})^2\sigmaz + \frac{\Delta_0}{p_F^2}\delta p_x \delta p_y \sigmax \}
\ee

where $v_F=p_F/m$, $v_\Delta=\Delta_0/p_F$. The linearized approximation consists of dropping the terms within brackets, in which case we obtain an anisotropic Dirac dispersion for the quasiparticles near the nodes:-
$$
E_L(p) = v_F \sqrt{(\delta p_x)^2 + \alpha^{-2}(\delta p_y)^2}
$$
where $\alpha =v_F/v_\Delta$, is the anisotropy of the Dirac dispersion. Following \cite{Simon&Lee} we estimate the temperature scale up to which this approximation can be trusted. We require that the value of the quadratic terms at the typical momenta of the thermally excited quasipartices be smaller than the linear contribution. For YBCO, this temperature turns out to be T$_{max} \sim$200Kelvin \cite{tmax}.

We now consider a similar low energy approximation for the {\it vortex state} of the d$_{xy}$ superconductor. First, we describe a proceedure to handle the non-single valued nature of the wave functions $\psi' = \left( \matrix{u' \cr v' }\right)$ of equation (\ref{psiprime}) \cite{Anderson}\cite{FT}, which is called the Franz-Tesanovic transformation, although we adopt a slightly different approach to its derivation here.

{\it Franz-Tesanovic Transformation:} Recall that we want to solve the eigenvalue equation;
\be
\label{Hprime}
H'_{BdG}\psi'= E \psi'
\ee
with the condition that $\psi'(x,y)$ is not single valued, but acquires a negative sign on circling an odd number of $\frac{hc}{2e}$ vortices. To implement this condition, we write $\psi'$ as a product of a fixed function that is multiple valued which precisely builds in the required sign changes ($\Phi(x,y)$), and a single valued wave function $\psi_{FT}(x,y)$. Thus,
\be
\label{psiFT}
\psi'(x,y) = \Phi(x,y)\psi_{FT}(x,y)
\ee

, and the preceeding eigenvalue problem can be formulated as an equivalent problem for the single valued wavefunction $\psi_{FT}$. In order that this is also a hermitian eigenvalue problem, we choose $\Phi(x,y)$ from the class of functions:
\be
\Phi_{\{q\}}(x,y) = \prod_i (\frac{z-z_i}{\bar{z}-\bar{z}_i})^{q_i/4}
\ee
where the $q_i$ are odd integers, and we have used the complex coordinates $z=x+iy$, $\bar{z}=x-iy$. The product runs over all $\frac{hc}{2e}$ vortices, located at $z_i=x_i+iy_i$. Note that this function is pure phase i.e. $|\Phi_{\{q\}}(x,y)|=1$ \cite{CS}. 
	The choice of the odd integers $q_i$ is arbitrary and represents a gauge degree of freedom. Clearly, physical results cannot depend on this choice, though in practice we may work with a particular set of $\{q_i\}$ that is convenient for calculation. Due to its singular nature, this transformation needs to be handled with some care especially in the linearized theory, a point which we will return to in Appendix B.

	Inserting (\ref{psiFT}) in (\ref{Hprime}), we have:
\begin{eqnarray}
\label{HFT}
H_{FT}\psi_{FT} &=& E \psi_{FT}\\
H_{FT} &=& \epsilon(\vec{p}+\vec{a}+\vec{P}_s\sigmaz)\sigmaz + \Delta(\vec{p}+\vec{a})\sigmax 
\end{eqnarray}
where $\vec{a}(x,y)$ is a real vector field given by:
\begin{eqnarray}
\label{a}
\vec{a} &=& -i\hbar \vec{\nabla} \log{\Phi_{\{q\}}(\vec{r})}\\
        &=& \frac\hbar2 \sum_i q_i \hat{z}\times \frac{\vec{r}-\vec{r}_i}{|\vec{r}-\vec{r}_i|^2} 
\end{eqnarray}
which implies
$$
\vec{\nabla}\times \vec{a}(\vec{r}) = \hbar\pi \sum_i q_i \delta^{(2)}(\vec{r}-\vec{r}_i)
$$
that solenoids of flux $\pi q_i$ of the $\vec{a}$ gauge field have been attached to the vortices. The sign change of the quasiparticles on circling a unit vortex is now accounted for by the Aharonov-Bohm effect arising from this solenoid of flux. Thus, a ficticious U(1) gauge field has been invoked to handle the (-1) phase factors acquired by a quasiparticle on circling a vortex. Clearly, this is a highly redundant description - in principle the U(1) gauge field $\vec{a}$ can account for any phase factor, which is here being restricted to just $\pm$1. The minimal choice that would take care of just these two factors, is an Ising (${\mathcal Z}_2$) gauge field, which however requires a real space lattice for its formulation\cite{Senthilfisher}.

{\it Linearization for the Vortex State:} Following \cite{Simon&Lee}, we consider low energy excitations near the nodal points, and neglect the effect of inter-node scattering. Then, we can expand the wavefunction as:
\begin{eqnarray}
\psi_{FT}(\vec{r}) &=& \mbox{e}^{ik_Fx}\psi_1(\vec{r})+\mbox{e}^{-ik_Fx}\psi_{\bar{1}}(\vec{r})\nonumber \\
	& & +\mbox{e}^{ik_Fy}\psi_2(\vec{r})+\mbox{e}^{-ik_Fy}\psi_{\bar{2}}(\vec{r})
\end{eqnarray}  
where the functions $\psi_i$ (i=$1$,$\bar{1}$,$2$,$\bar{2}$) are considered to be slowly varying on the scale of $k_F^{-1}$. The problem then reduces to solving, at each node:
\be
(H^{FT}_i + \Delta H)\psi_i(\vec{r}) = E \psi_i(\vec{r})
\ee
where the $H^{FT}_i$ represent the linearized part of the Hamiltonian (\ref{HFT}):
\begin{eqnarray}
H^{FT}_1 &=& -H^{FT}_{\bar{1}} \\ \nonumber
	&=& v_F\{(p_x + a_x)\sigmaz + \alpher (p_y+a_y)\sigmax+\hat{x}\cdot\vec{P}_s(\vec{r})\}\\
H^{FT}_2 &=& -H^{FT}_{\bar{2}} \\ \nonumber
	&=& v_F\{(p_y + a_y)\sigmaz + \alpher (p_x+a_x)\sigmax+\hat{y}\cdot\vec{P}_s(\vec{r})\}
\end{eqnarray}
with $\vec{p}=-i\vec{\nabla}$. The remaining part of the Hamiltonian, $\Delta H$, arises from the curvature of the electon dispersion and of the gap function and is given by:

\begin{eqnarray}
\Delta  H &=& \frac1{2m} [(\vec{p}+\vec{a})^2 + \vec{P}_s^2(\vec{r})]\sigmaz +  \frac1{2m} \{\vec{p}+\vec{a}, \vec{P}_s(\vec{r}) \}\One \nonumber \\
	& & + \frac{\Delta_0}{2p_F^2}\{ p_1 + a_1, p_2 + a_2\}\sigmax
\label{DeltaH_FT}
\end{eqnarray}
where $\{,\}$ denotes antisymmetrization; $\{A,B\}=AB + BA$.

It may now be argued that the linear piece $H^{FT}_i$, dominates over the curvature terms $\Delta H$ for magnetic fields much smaller than $H_{c2}$. Let us denote the typical separation between vortices by $d \propto 1/\sqrt{B}$. While the linear part of the Hamiltonian has terms of order $E_1 = \hbar v_F/d$ (50K Tesla$^{-\frac12}$ in YBCO) or of order $E_2=\hbar v_\Delta/d$ (3K Tesla$^{-\frac12}$ in YBCO), the largest terms in $\Delta H$ are of order $E_1^2/E_F$ (0.5K Tesla$^{-1}$ in YBCO), so long as we assume that the quasiparticle trajectories do not come too close to the vortex core. Thus, in the magnetic field range of interest of a few Tesla, we can at a first approximation neglect the curvature terms $\Delta H$ and simply solve the linearized problem. Subsequently we will include the effect of these curvature terms (they are crucial to producing a finite thermal Hall signal), and verify that their effect is indeed a perturbation on the linearized problem.  

	We are therefore led to consider the linearized problem, say at node 1:
\begin{eqnarray}
H^{FT}_1 \psi &=& E\psi \nonumber \\
H^{FT}_1 &=& v_F\{(p_x + a_x)\sigmaz+(p_y+a_y)\sigmax + P_{sx}(\vec{r})\One\}
\end{eqnarray}
We have already noted, that in addition to the interaction with the amplitude variation of the gap (which has been neglected) the quasiparticles interact with the vortices via the superflow $\vec{P}_s$ and the Berry phase of (-1) acquired on circling a fundamental vortex. For the sake of clarity, it is helpful to first consider a situation where the Berry phases are inactive, and the quasiparticles only see the superflow. This is formally accomplished by considering the case of a vortex lattice of $\frac{hc}e$ (double) vortices. Then, there are no nontrivial Berry phase factors and hence the $\vec{a}$ fields (which were inducted to keep track of these phase factors), may be dropped. Equivalently, it may be noted that for the case of $\frac{hc}e$  vortices, the transformation (\ref{U}) {\it is} single valued. Below, we will analyze the linearized problem in the presence of a $\frac{hc}e$ vortex lattice. Armed with this understanding, we then consider the physically relevant case of a lattice of $\frac{hc}{2e}$ vortices. Although our conclusions from the $\frac{hc}e$ vortex lattice case go over unchanged, the reasoning there is a little more involved.

\subsection{Vortex Lattice of $\frac{hc}e$ vortices in the linearized approximation:}
Consider the linearized hamiltonian at node 1, for this conceptually simpler case of a vortex lattice of  $\frac{hc}e$ vortices. 
\be
\label{H1linear}
H_1 = v_F\{p_x\sigmaz + \alpher p_y\sigmax + P_{sx}(\vec{r})\One\}
\ee
where the superflow $\vec{P}_s=(P_{sx},P_{sy})$ is given by the gauge invariant combination, 
$$
\vec{P}_s = [\frac12 \vec{\nabla}\phi - e\vec{A}]
$$
and satisfies,
\be
\label{curlps2}
\vec{\nabla}\times \vec{P}_s = [(h\sum_i \delta^{2}(\vec{r}-\vec{R}_i))-eB(\vec{r})]\hat{z}
\ee
where the sum runs over all the vortices in the lattice, at positions $\vec{R}_i$. This differs from equation (\ref{curlps}) in that the strength of the vortices is now doubled. For a periodic lattice of vortices, $\vec{P}_s$, being a physical quantity is also a periodic function with the same period as the vortex lattice. Formally, this can be seen by noting that:-
\be
\int_{\mbox{unit cell}} d^2r (\vec{\nabla}\times \vec{P}_s) = 0
\ee
 from flux quantization. Thus,
$$
\oint _{{\mathcal B}} \vec{P}_s \cdot d\vec{l} =0
$$
on the boundary ${\mathcal B}$ of the unit cell, which is consistent with a periodic superflow $\vec{P}_s$. 

	Consequently, the Hamiltonian $H_1$ (\ref{H1linear}) is that of a Dirac particle in the presence of a periodic scalar potential (played by $P_{sx}(\vec{r})$). Cearly, a band structure will result, and the eigenstates can be labelled by the band index, and the crystal momentum $\vec{k}$ that takes values within the Brillouin zone. What follows is a symmetry analysis of the spectrum of $H_1$. One of the main issues we address is whether the Dirac node at zero energy, that is present for the free Dirac Hamiltonian (and corresponds to the node of a d$_{xy}$ superconductor in the pure state) survives in the presence of the periodic superflow term $P_{sx}(\vec{r})$. Our results are as follows. Dirac nodes (band touchings) survive in the spectrum of $H_1$ at those points in the Brilloin zone that are invariant under the $\vec{k}\rightarrow -\vec{k}$ transformation. The symmetry that protects these nodes, which we call ${\mathcal T}_{Dirac}$, is obtained as a consequence of the linearization. Further, we find that there is a Dirac node centered at {\it zero energy} if the vortex lattice posesses inversion symmetry.

\section{Dirac Nodes in the Linearized Problem:}
	We begin by analyzing the conceptually simpler case of a vortex lattice of $hc/e$ vortices, where the Berry phase factors are absent, and we have only the coupling of the quasiparticles to the superflow to deal with. We then tackle the $hc/2e$ vortex lattice case along parallel lines.

\subsection{Vortex Lattice of $\frac{hc}{e}$ Vortices}

\begin{figure}
\epsfxsize=1.4in
\centerline{\epsffile{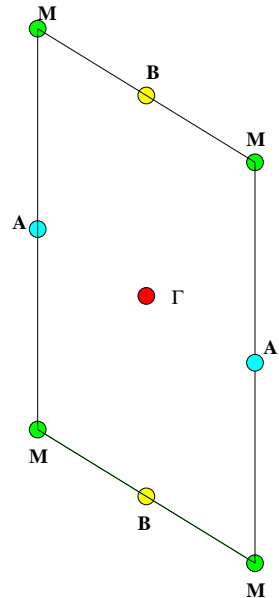}}
\vspace{0.1in}
\caption{The Brillouin zone for an oblique lattice of vortices. The points of the Brillouin zone that are taken to themselves under time reversal symmetry are shown as $\Gamma$, {\bf A,B} and {\bf M} points. At these points degenerate doublets and hence Dirac cones are expected within the linearised theory for all lattices of $\frac{hc}e$ vortices. The same holds for lattices of $\frac{hc}{2e}$ vortices.}
\label{fig1}
\end{figure}

We will consider the linearized Hamiltonian for node 1 :
$$
H_1 = v_F\{p_x\sigmaz + \alpher p_y\sigmax + P_{sx}(\vec{r})\One\}
$$
which is equation (\ref{H1linear}) of the previous section, where $\vec{P}_s(\vec{r})$ for an arbitrary lattice of vortices, is a periodic function with the same period as the lattice and satisfies equation (\ref{curlps2}). This gives rise to a band structure for the quasiparticles, and the eigenstates are labelled by a band index and a crystal momentum ($\vec{k}$), which takes values within the Brillouin zone. We consider for generality an oblique vortex lattice, which posess no additional symmetries. The Brillouin zone for such a vortex lattice is depicted in Figure \ref{fig1}. 

{\it A Symmetry of the Linearized Hamiltonian:} Consider an eigenstate $\psi_{\vec{k}}$ of the linearized Hamiltonian $H_1$ with eigenvalue $E$ and crystal momentum $\vec{k}$: 
$$
H_1 \psi_k = E \psi_k
$$

Then the transformed wavefunction $\phi_{-\vec{k}} = \TAU \psi^*_{\vec{k}}$ is also an eigenstate with energy $E$ but crystal momentum $-\vec{k}$, where $\TAU$ is the dimension two antisymmetric matrix:
$$
\TAU =  \left( 
	\matrix{
		0 & 1\cr
		-1 & 0} 
	\right)
$$
This can be verified by noting that the linearized Hamiltonian is taken to itself under the transformation:
\be
-\TAU H_1^* \TAU = H_1
\ee
where we have used the identities

\begin{eqnarray*}
\TAU^2 &=& -{\mathbf{1}}\\ 
\TAU \mathbf{\sigma}_i^* \TAU &=& \mathbf{\sigma}_i
\end{eqnarray*}
Since this transformation is formally equivalent to the time reversal operation for {\it Dirac} particles, we will call it ${\mathcal T}_{Dirac}$, although, as explained below, it is distinct from the physical time reversal transformation for this problem. This symmetry ensures that states with crystal momentum $\vec{k}$ and $-\vec{k}$ have the same energy. For those points in the Brillouin Zone that are taken to themselves  under time reversal (modulo  a reciprocal lattice vector: $\vec{k} \equiv -\vec{k}$ [mod $\vec{G}$)]) there is a degenerate pair of states $\psi_{\vec{k}}$ and $\TAU \psi^*_{\vec{k}}$. These states are orthogonal, from  the antisymmetry of $\TAU$. Hence we find a degenerate doublet at these special points in the Brillouin Zone, which are shown in Figure \ref{fig1} as the $\Gamma$ ($\vec{k}=0$), A, B and M points. The spectrum at these special points of the Brillouin Zone is composed entirely of degenerate pairs, and as we will see shortly, on moving a little bit away from these points in crystal momentum, the states split and give rise to the energy dispersion of a massless Dirac particle.

	The symmetry operation ${\mathcal T}_{Dirac}$ that operates on the quasiparticle excitations at a single node is {\it distinct} from the physical time reversal operation that would transform states at one node into states at the opposite node. Rather, ${\mathcal T}_{Dirac}$ is obtained as a symmetry as a consequence of linearizing the electron dispersion, and it is easily seen that the curvature terms, such as for example $\frac1{2m} p^2\sigmaz$, violate this symmetry. Thus, although ${\mathcal T}_{Dirac}$ is not a symmetry of the entire problem, since the curvature terms that violate it are so small, it is still a good approximate symmetry. For the linearized theory of course, it is an exact symmetry.

{\it Dirac Cones from Degenerate Doublets:} As we move away from the special points in the Brillouin zone at which degenerate doublets are found, the crystal momentum splits these states and gives rise to a Dirac cone dispersion. This is most easily appreciated by analogy with the pure case. There, the free Dirac Hamiltonian, $H_{free}=v_F p_x \sigma_z + v_\Delta p_y \sigma_x$, has a degenerate pair of states at the center of the Dirac cone (at $\vec{p}=0$) given by the two constant spinors that are eigenstates with zero energy. Moving away in momentum, which is exactly analogous to turning up the Zeeman field on a spin one-half particle, the states are split into $\pm$E pairs, with the splitting being linearly proportional to the momentum. This give rise to the massless Dirac dispersion that we expect for this Hamiltonian. In a similar way, we find here that Dirac cones arise centered at the special points at which the degenerate doublets are present.

Consider a pair of degenerate states, $\psi(\vec{r})$ and $\TAU\psi^*(\vec{r})$, at one of these special points in the Brillouin zone. The effect of moving away from this point by crystal momentum $\delta \vec{k}$ can be accounted for by adding the piece:
\be
\label{deltaHk} 
\delta H _{\delta k} =   v_F \delta k_x \sigmaz + v_\Delta \delta k_y \sigmax
\ee
to the Hamiltonian and leaving unchanged the boundary condition for the wavefunction. If the deviation in crystal momentum is small (compared to the reciprocal lattice vectors), this additional piece will barely mix the different pairs of energy eigenstates, and so can be treated in degenerate perturbation theory within the two dimensional subspace of the degenerate doublet. The perturbation, projected into the subspace of  $\psi(\vec{r})$ and $\TAU\psi^*(\vec{r})$, takes the form:

\be
[\delta H_{\delta k}]_{\mbox{Proj}}=  \left( 
			\matrix{
			\int_{\vec{r}}\psi^\dag  \delta H_{\delta k} \psi & -\int_{\vec{r}}\psi^T \TAU \delta H_{\delta k} \psi \cr
			\int_{\vec{r}}\psi^\dag  \delta H_{\delta k}\TAU \psi^* & -\int_{\vec{r}}\psi^\dag \delta H_{\delta k} \psi} 
			\right)
\ee
where the integral $\int_{\vec{r}}$ runs over the unit cell and we have used the notation $\psi^\dag(\vec{r}) = \psi^{*T}(\vec{r})$. Clearly, an explicit knowledge of the wavefunctions is needed to diagonalize this Hamiltonian, but a few general observations can be made right away. First, the eigenvalues obtained here are the energy splitting ($\delta E_{\delta k}$) of the previously degenerate pair of states, which will appear in a plus-minus pair since the projected Hamiltonian has vanishing trace. Also, if we denote by $\theta_{\delta k}$ the angle the vector $\vec{\delta k}$ makes with the $k_x$ axis, the energy splitting may be written as:
\be
\delta E_{\delta k} = \pm v_F A(\theta_{\delta k}) |\vec{\delta k}|
\ee

which is a massless Dirac particle with an anisotropic dispersion. Calculating the anisotropy function $A(\theta_{\delta k})$ requires an explicit knowledge of the wavefunctions.

	If the vortex lattice posesses reflection symmetry about the y axis (x axis), then the dispersion of the Dirac nodes of the linearized Hamiltonian at node 1 (2) takes a particularly simple form. The dispersion can then be written as:
\be
\delta E_{\delta k} = \pm \sqrt{(v'_F)^2 \delta k_x^2 + (v'_\Delta)^2 \delta k_y^2 }
\ee 
where again an explicit knowledge of the wavefunctions is required to compute the renormalized velocities $v'_F$ and $v'_\Delta$. The effect of the renormalization can be substantial and therefore the energy scale at which the Dirac node may be expected to make its presence felt can be quite different from $E_2$, as one might naively expect, as pointed out in \cite{Marinelli}.

\begin{figure}
\epsfxsize=2.8in
\centerline{\epsffile{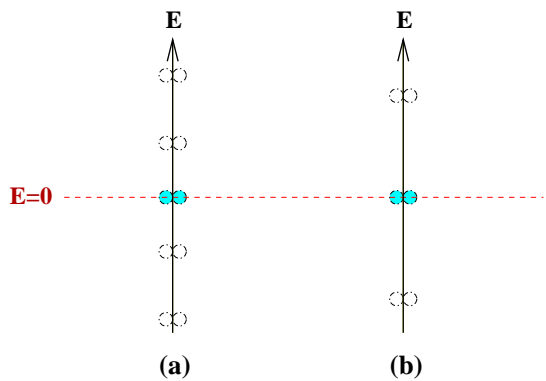}}
\vspace{0.1in}
\caption{Dirac node at zero energy for inversion symmetric lattices: (a) Energy spectrum of a free Dirac particle on a torus ($\Gamma$ point). There are an odd number of degenerate pairs of energy states, due to particle-hole symmetry and the pair at zero energy. (b) In the presence of the linearized Doppler term ($P_{sx}$) for inversion symmetric lattices of $\frac{hc}{e}$ vortices, particle-hole symmetry of the spectrum is preserved, and the degenerate pairs of states cannot split due to symmetry under ${\mathcal T}_{Dirac}$. To keep the count of energy levels the same as for the free case, there has to be a pair of states at {\em E}=0. Similar considerations apply for the case of inversion symmetric lattices of physical $\frac{hc}{2e}$ vortices.}
\label{fig2}
\end{figure}

{\it Inversion Symmetry and the Dirac Node at Zero Energy:} If the vortex lattice posesses inversion symmetry, that is, if it is invariant under the transformation $\vec{r}\rightarrow -\vec{r}$ (assuming the origin is the center of inversion), then it is easy to see from (\ref{curlps2}) that the superflow $\vec{P}_s$  satisfies:
\be
\vec{P}_s(-\vec{r}) = -\vec{P}_s(\vec{r})
\ee 
. This leads to a particle hole symmetry of the linearized Hamiltonian. If $\psi (x,y)$ is an  eigenstate of the Hamiltonian $H_1$ (\ref{H1linear}) with energy $E$, then   $\psi (-x,-y)$ is also an eigenstate but with energy $-E$ \cite{ptclhole}.  
	Inversion symmetry ensures there is a degenerate doublet of the linearized Hamiltonian $H_1$ at the $\Gamma$ ($\vec{k}=0$) point at {\it zero energy}. The argument is as follows - let us focus on the spectrum at the $\Gamma$ point; in which case we need to solve for the eigenstates of $H_1$ on the unit cell with periodic boundary conditions - i.e. on a torus. First, consider the case without the Doppler term, that is a free Dirac particle on a torus:
$$
H_{free}=v_F p_x \sigmaz + v_\Delta p_y \sigmax
$$
which can easily be solved. Clearly, this has a pair of states at zero energy, given by the product of the constant solution times any spinor. The rest of the states also occur in degenerate pairs, and for every pair of states at energy $E\neq0$ there is a pair of states at energy $-E$; a consequence of the free Dirac Hamiltonian respecting the ${\mathcal T}_{Dirac}$ and inversion symmetries. This spectrum is sketched in figure \ref{fig2}(a). Thus, in the free case, the spectrum consists of an `odd' number of degenerate pairs, due to the existence of the pair at zero energy (this can be made more rigorous by introducing an ultraviolet cutoff, and hence a finite number of states). Now, turning on the Doppler term $P_{sx}$ for inversion symmetric vortex lattices, preserves both the ${\mathcal T}_{Dirac}$ as well as particle-hole symmetry. Therefore we still have states coming as degenerate doublets, in a particle-hole symmetric spectrum. Since the total number of pairs of states cannot change from the free case, we are forced to have a degenerate doublet at {\it zero energy}, so that the total number of pairs of degenerate states remains odd, as it was for the free case \cite{Odd}. This argument is illustrated in Figure \ref{fig2}.

	Notice that this argument is also valid perturbatively. If we imagine continuously turning up from zero the value of the Doppler term, the zero energy doublet that is present for the free Dirac Hamiltonian can neither split (it is protected by ${\mathcal T}_{Dirac}$) nor move away from zero energy (thanks to particle-hole symmetry) and hence continues to exist at zero energy even for the Hamiltonian $H_1$ that includes the superflow of an inversion symmetric vortex lattice. This can also be verified to all orders in perturbation theory, the details of which may be found in Appendix A.

	This doublet of states at zero energy for inversion symmetric lattices, will give rise to a Dirac cone centered at zero energy, by our previous arguments. Thus, in this sitation we have been able to access the nature of the low energy physics solely via the use of symmetry arguments.

\subsection{Vortex Lattice of $hc/2e$ Vortices}
\label{sect3}
	We now turn to the physically relevant case of a vortex lattice of $hc/2e$ vortices within the linearized approximation. For convenience, we collect here the relevant formulae derived in the previous section; the linearized Hamiltonian at node 1 is:
$$
H^{FT}_1 = v_F\{(p_x + a_x)\sigmaz+(p_y+a_y)\sigmax + P_{sx}(\vec{r})\One\}
$$
where the $\vec{a}$ fields implement the Berry phase factor for circling a vortex which we have taken to be:
\begin{eqnarray*}
\vec{a} &=& -i\hbar \vec{\nabla} \log{\Phi_{\{q\}}(\vec{r})}\\
\Phi_{\{q\}}(z,\bar{z}) &=& \prod_i (\frac{z-z_i}{\bar{z}-\bar{z}_i})^{q_i/4}
\end{eqnarray*}
where the $q_i$ are arbitrary odd integers. More generally, the $\vec{a}$ fields just need to satisfy:
\be
\vec{\nabla}\times \vec{a}(\vec{r}) = h\pi \sum_i q_i \delta^{(2)}(\vec{r}-\vec{r}_i)
\label{curla}
\ee

which attaches solenoids of $\pi q_i$ ficticious flux to the vortices. Clearly, physical results should not depend on the choice of $q_i$. However, since we have a vortex lattice, it will be convenient to pick a set of $q_i$ that give rise to a periodic $\vec{a}$ field. Imagine choosing a set of $q_i$, so that the unit cell for the problem now contains {\em n} vortices. Periodicity of the vector potential $\vec{a}$ requires that the total flux corresponding to this vector potential vanishes over the unit cell, i.e.:
\be
\sum_{\alpha=1}^n q_\alpha = 0 
\ee

Where the sum is over the vortices in a unit cell. Since the $q_\alpha$ are odd integers, this requires that the number of vortices in the unit cell, {\em n}, be an even number. Therefore, a vortex lattice that has one physical vortex per unit cell, will require at least a doubling of the unit cell when considered in this way. Once again, the eigenfunctions can be labelled by crystal momenta $\vec{k}$ that takes values in the appropriate Brillouin zone which is now determined, not just by the periodicity of the vortex lattice, but also by the choice of  $q_i$s.

{\it Invariance Under Combined ${\mathcal T}_{Dirac}$ and Gauge Transformation:} It is easily seen that the Hamiltonian $H^{FT}_1$ is not invariant under ${\mathcal T}_{Dirac}$; i.e. under the transformation $\psi \rightarrow \tau \psi^*$, the sign of the vector potential $\vec{a}$ is inverted. However, since the flux associated with the $\vec{a}$ field is exactly $\pi$ (times an odd integer), it is possible to arrange for a gauge transformation that would reverse its sign, which when used in combination with ${\mathcal T}_{Dirac}$, would leave the Hamiltonian invariant. Thus, if $\psi_{\vec{k}}(\vec{r})$ is an eigenfunction of $H^{FT}_1$ with energy {\em E} and crystal momentum $\vec{k}$, then
\be
\label{combination}
\phi_{-\vec{k}}(\vec{r}) = \Phi^2_{\{q\}}(\vec{r}) \TAU \psi_{\vec{k}}^*(\vec{r})
\ee
is also an eigenfunction of the Hamiltonian with energy {\em E}. The factor of $\Phi^2_{\{q\}}$  performs a gauge transformation that invert the sign of the gauge fields $\vec{a}$. It is crucial that $\Phi^2(\vec{r})$ is a single valued function, so this gauge transformation is only allowed because we have the special case of $\pi$ flux solenoids. Also, it is easily shown that $\Phi^2(\vec{r})$ is a periodic function, so the wavefunction $\phi_{-\vec{k}}(\vec{r})$ carries crystal momentum $-\vec{k}$. Thus, once again, states with crystal momentum $\vec{k}$ and  $-\vec{k}$ have the same energy. At those points in the Brillouin zone that are taken to themselves  under time reversal, modulo  a reciprocal lattice vector, [$\vec{k} \equiv -\vec{k}$ (mod $\vec{G}$)] the energy levels appear as degenerate doublets. This follows from the fact that the two wavefunctions ($\psi_{\vec{k}}(\vec{r})$ and $\Phi^2_{\{q\}}\TAU \psi_{\vec{k}}^*(\vec{r})$) are always othogonal from the antisymmetry of $\TAU$. These degenerate doublets will lead to Dirac cones in their vicinity by the same argument as before. Hence, for any vortex lattice we expect the spectrum of  $H^{FT}_1$ to posess Dirac cones (band touchings). 

As we have noted, we are only allowed to make the gauge transformation above because we are dealing with $(2m+1)\pi$ solenoid fluxes of the ficticious gauge field $\vec{a}$, that leads to $\Phi^2_{\{q\}}(\vec{r})$ being a single valued function. For a general value of the ficticious flux, time reversal symmetry of the Dirac equation is broken in an essential way and cannot be fixed by such a gauge transformation. This also implies that in general, the doubly degenerate states that we obtain are not accessible within perturbation theory starting with a free Dirac equation. In doing perturbation theory we are implicitly assuming that we can gradually crank up from zero the value of the ficticious flux - however since the doublet only appears at the $\pi$ flux values, it is missed in perturbation theory (unless some additional symmetry, that preserves the degenerate doublets for all values of the ficticious flux, is present). This issue is discussed in more detail in Appendix A, and some subtle aspects of the gauge transformation that we have just used are considered in Appendix B.

\begin{figure}
\epsfxsize=3.2in
\centerline{\epsffile{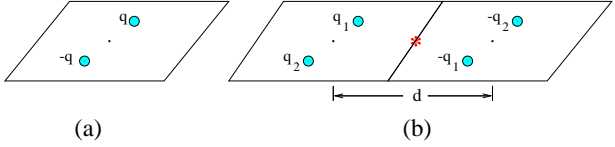}}
\vspace{0.1in}
\caption{Different F-T gauge choices ($q_i$) for a vortex lattice of two physical vortices per unit cell. (a) Smallest unit cell possible, under inversion the $q$ and $-q$ vortices are interchanged. (b) Doubling of the unit cell due to gauge choice. Once again the ficticious gauge field flux has been chosen, for convenience, to map onto its negative under inversion about the origin (*).}
\label{fig3}
\end{figure}

{\it Inversion Symmetry and the Dirac Cone at Zero Energy:} Consider a vortex lattice that posesses inversion symmetry (about the origin, say) i.e. it is invariant under the transformation $\vec{r}\rightarrow -\vec{r}$. We argue here that in this case there exists a degenerate doublet, and hence a Dirac cone centered {\it at zero energy}. 

For such a lattice it is always possible to make a gauge choice of the set \{$q_i$\} for which, under inversion, $q_i \rightarrow -q_i$. A couple of examples are shown in Figure \ref{fig3}. In case we begin with a set of $q_i$ that do not satisfy this condition,  one can always make a gauge transformation that does not alter any of the physics but brings the $q_i$s into this form, which happens to be convenient for the following analysis. In this gauge we have that the $\vec{a}$ fields are even under inversion, i.e.
\begin{equation}
\vec{a}(-x,-y) = \vec{a}(x,y)
\end{equation} 
which is easily seen from equation (\ref{curla}) for the curl of $\vec{a}$ and the fact that the position of a $q$ vortex is taken by a $-q$ vortex under inversion. Of course we still retain the fact that the superflow is odd under inversion:
\be
\vec{p}_s(-x,-y) = -\vec{p}_s(x,y)
\ee

These facts imply a particle-hole symmetry for the Hamiltonian $H^{FT}_1$. If $\psi(x,y)$ is an eigenfunction of $H^{FT}_1$ with energy {\em E}, then $\tau \psi^*(-x,-y)$ is also an eigenfunction but with energy -{\em E} \cite{Marinelli}.

Inversion symmetry ensures that the degenerate doublet of Hamiltonian $H^{FT}_1$ at the $\vec{k}$=0 ($\Gamma$) point is at zero energy. The argument is as follows - for the $\Gamma$ point we need to solve for the eigenstates of Hamiltonian $H^{FT}_1$ on a torus. First consider the case without the Doppler term and the $\vec{a}$ field, that is,  a free Dirac particle on a torus. 
$$
H_{free} = v_F \{p_x \sigmaz + \alpher p_y \sigmax \}
$$
This has a pair of states at zero energy, given by the constant solution and any spinor. The rest of the states also occur in pairs (due to ${\mathcal T}_{Dirac}$) and for every pair at energy {\em E} $\not= 0$ there is a pair of states at energy -{\em E} (from particle-hole symmetry). Thus in the free case, as sketched in Figure \ref{fig2}(a), there are an `odd' number of degenerate doublets due to the pair at zero energy (this can be made more rigorous by introducing an ultraviolet cutoff, and hence a finite number of states). Now, turning on the Doppler term and the gauge field $\vec{a}$, the energy levels will again appear as degenerate pairs, from symmetry under the combined effect of ${\mathcal T}_{Dirac}$ and a gauge transformation, as discussed earlier. For the case of inversion symmetric vortex lattices, particle-hole symmetry is preserved as well. Therefore we have the degenerate doublets appearing in a particle-hole symmetric fashion. Since the total number of pairs of states cannot change from the free case, where we had an odd number of pairs, we are forced to have a degenerate doublet at {\it zero energy}, as illustrated in Figure \ref{fig2}(b). The doublet of states at zero energy will give rise to a Dirac cone centered at {\em E}=0 by our previous arguments.   Notice that this is a non-perturbative argument for the existence of a zero energy pair of states in inversion symmetric lattices of $\frac{hc}{2e}$ vortices. In general it is not possible to access these states within perturbation theory since they only appear when the flux of the gauge field $\vec{a}$ takes on values that are odd multiples of $\pi$. An alternate argument for the existence of this Dirac node at zero energy for inversion symmetric $\frac{hc}{2e}$ vortex lattices, that does not invoke the F-T transformation, is provided in Appendix B. 

	Thus, for the case of an inversion symmetric lattice of $hc/2e$ vortices as well, the linearized theory predicts the existence of a Dirac node at zero energy.

\section{Beyond the Linearized Approximation: Massive Dirac Quasiparticles and Quantized Thermal Hall Effect}

We now consider the effect of the subdominant curvature terms
($\Delta H$) on the spectrum obtained from the linearized equations.
In view of the smallness of these terms  compared to
the linearized Hamiltonian, their primary effect will be to lift
degeneracies that are present in the spectrum of the linearized
problem. Therefore we study the effect of these terms
near the Dirac cones (band touchings) of the linearized problem. In order to simplify the discussion we shall consider the the situation of a vortex lattice of \hcbye (double) vortices. For the physically relevant case of the
\hcbytwoe vortex lattice, the discussions runs on very similar lines,
although $\Delta H$ in that case involves more terms arising
from the Franz-Tesanovic gauge field $\vec{a}$ (\ref{DeltaH_FT}). We do not present the details of that case here.
\begin{figure}
\epsfxsize=3.2in
\centerline{\epsffile{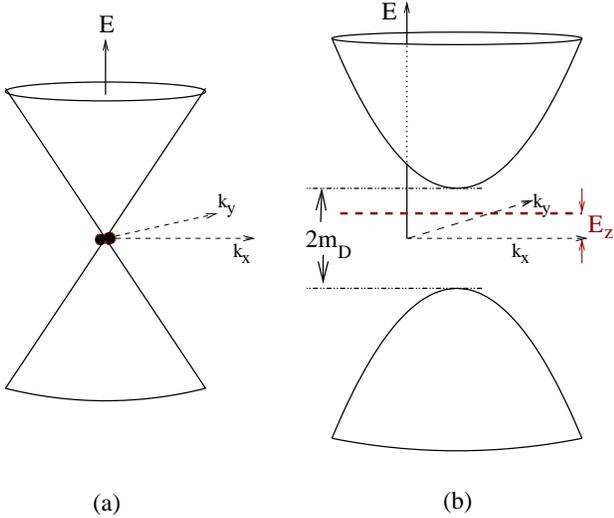}}
\vspace{0.1in}
\caption{On including the effect of curvature terms, the Dirac cone obtained at zero energy for inversion symmetric lattices (a) now acquire a small mass gap $2m_D$ (b). The chemical potential for quasiparticles, that is set by the Zeeman energy $E_Z$ is shown here as lying within the gap.}
\label{gapping}
\end{figure}
	
\subsection{Massive Dirac Quasiparticles}		

The problem at hand then is to consider the effect of:
\be
\Delta H = \frac1{2m} [p^2 + \Ps^2(\vec{r})]\sigmaz +
\frac1{2m}\{p,P_s(\r) \} {\mathbf{1}} + \frac{\Delta_0}{p_F^2}p_xp_y
\sigmax 
\label{DeltaH}
\ee

the curvature terms, on the spectrum of the linearized Hamiltonian at, say,  node 1, 
$$ 
H_1 = v_F\{p_x \sigmaz + \alpher p_y \sigmax + \Psx {\mathbf{1}}\}
$$ 

We  have seen  that the band structure of the linearized problem posesses Dirac cones at special points in the Brillouin zone where there is  a pair of degenerate states. This degeneracy is a
result of the  invariance  of    the linearized Hamiltonian      under
the symmetry operation \Tdirac.  The  curvature terms however   do not respect  this symmetry
(under   \Tdirac,  $\Delta  H$ changes  sign) and so are expected to split the
degenerate doublets, giving rise to a gap  in the dispersion  (see
Figure   \ref{gapping}).  This may be analyzed within degenerate perturbation theory, the details of which are below. 

Consider a  degenerate
doublet ($\psi$, $\TAU \psi^*$) of the linearized Hamiltonian at one of
the special points in  the Brillouin zone. We  may treat the effect
of $\Delta H$ within  degenerate perturbation theory, since  the other
energy  levels at this crystal momentum  are separated by energies of
order  E$_1$ or   E$_2$, that are much  larger  than   the  strength of    the
perturbation   $\Delta H$.  On  projecting  $\Delta  H$ into this  two
dimensional subspace, it can be written as:
\be
\label{ndeltaH}
[\Delta H]_{Proj} = (\frac{\hbar eB}{m})\vec{n}_{\Delta H}\cdot \mbox{\boldmath{$\vec{\tau}$}}
\ee
where  {\boldmath{$\vec{\tau}$}} are the  Pauli matrices that act within this two dimensional subspace, and there is
no term proportional  to  {\boldmath{$1$}}, since $\Delta  H$  changes
sign  under \Tdirac. This defines for us the vector $\vec{n}_{\Delta H}$. The   dependence  on the  magnetic field,   B, is
explicitly exhibited in the prefactor. Further, if we neglect the last
term in $\Delta  H$  (\ref{DeltaH}), that arises from the curvature of the gap and is smaller  than the other
terms by a factor of  $\frac{\Delta}{E_F}$, then $\vec{n}_{\Delta  H}$
defined above is only a function  of the anisotoropy $\alpha$, and the
vortex  lattice geometry - i.e.  the  type of lattice and  its orientation relative to the nodes.

We now derive   the effect of the perturbation $[\Delta H]_{Proj}$  on the dispersion around the special Brillouin zone points. Consider making a small excursion in crystal momentum $\vec{k}$ ($|\vec{k}| \ll 2\pi/d$ the typical reciprocal lattice vector) from this point. As in (\ref{deltaHk}), we obtain an additional term in the linearized Hamiltonian $\delta H_{\vec{k}} = k_x \sigmaz + \alpher k_y \sigmax$ which when projected into the two dimensional space of the degenerate doublet takes the form
\be
[\delta H_{\vec{k}}]_{Proj} = v_F [k_x \vec{n}_{k_x}\cdot \vec{\TAU}+ \alpher k_y \vec{n}_{k_y}\cdot \vec{\TAU}]
\label{nk}
\ee
where, once again there is no term proportional to {\boldmath{$1$}}, since $\delta  H_{\vec{k}}$  changes sign  under \Tdirac. The $\vec{n}_{k_x}$ and $\vec{n}_{k_y}$ are, in general, a pair of linearly independent vectors, that are defined by the above equations. It may easily be seen that the dispersion resulting from the sum of these projected Hamiltonians $[\Delta H]_{Proj}+[\delta H_{\vec{k}}]_{Proj}$, is that of a massive Dirac particle with mass 
\be
\label{m_D}
m_D = (\frac{\hbar eB}{m})\vec{n}_{\Delta H} \cdot [\frac{\vec{n}_{k_x}\times \vec{n}_{k_y}}{|\vec{n}_{k_x}\times \vec{n}_{k_y}|}]
\ee
i.e. the component of $\vec{n}_{\Delta H}$ perpendicular to the plane defined by $\vec{n}_{k_x}$ and $\vec{n}_{k_y}$. This is most easily seen for the case when $\vec{n}_{k_x}$ and $\vec{n}_{k_y}$ are orthogonal \cite{orthogonal}. Then, if $\hat{n}_{k_x}$ ($\hat{n}_{k_y}$) is the unit vector in the direction of $\vec{n}_{k_x}$ ($\vec{n}_{k_y}$) we can write:
\begin{eqnarray}
\label{orthogonal case}
[\Delta H + \delta H_{\vec{k}}]_{Proj} &=& v_F'(k_x - k^0_x)\hat{n}_{k_x}\cdot \vec{\TAU} + v_\Delta '(k_y - k^0_y)\hat{n}_{k_y}\cdot \vec{\TAU} \nonumber \\
	& & + m_D (\hat{n}_{k_x}\times \hat{n}_{k_y})\cdot \vec{\TAU}
\end{eqnarray}
. The center of the dispersion is now at ($k^0_x$, $k^0_y$) = -$(\frac{\hbar eB}{m v_F})(\vec{n}_{\Delta H}\cdot \hat{n}_{k_x}, \alpha \vec{n}_{\Delta H}\cdot \hat{n}_{k_y})$ and the renormalized velocities are $v'_F=v_F|\vec{n}_{k_x}|$ and $v'_\Delta=v_\Delta|\vec{n}_{k_y}|$ and the mass term $m_D = (\frac{\hbar eB}{m})\vec{n}_{\Delta H} \cdot (\hat{n}_{k_x} \times \hat{n}_{k_y}) $. In an appropriately chosen basis the above projected Hamiltonian will take the form:
\begin{eqnarray}
[\Delta H + \delta H_{\vec{k}}]_{Proj} &=& v'_F(k_x - k^0_x)\TAU_1 + v'_\Delta (k_y - k^0_y)\TAU_2 \nonumber \\ 
	& & + m_D \TAU_3
\end{eqnarray}
. Clearly, the resulting dispersion is:
\be
E(\vec{k})=\pm \sqrt{[v'_F(k_x - k^0_x)]^2 +[v'_\Delta (k_y - k^0_y)]^2 + m_D^2}
\ee
 i.e. that of a massive Dirac particle with mass $m_D$, centered at the crystal momentum ($k^0_x$, $k^0_y$) as shown in Figure \ref{gapping}. Of course, to calculate these quantities requires an explicit knowledge of the wavefunction $\psi$. Thus, the Dirac nodes obtained within the linearized approximation acquire small gaps on inclusion of the curvature terms. An explicit numerical calculation of these gaps is presented later in this section. Below, we invoke some well known properties of massive Dirac particles in two dimensions, to derive some consequences for quasiparticle transport in the vortex lattice state.

\subsection{Quantized Thermal Transport}

	The low temperature transport properties of  quasiparticles
	depends strongly on the nature of the spectrum at the chemical
	potential. For vortex lattices that posess inversion symmetry,
	the particle-hole symmetry of  the spectrum at each  node will
	allow  us  to make some precise  statements  regarding the low
	temperature transport properties -  therefore in what  follows
	we specialise to the case of inversion symmetric lattices.

	For   inversion  symmetric  lattices,   within the  linearized
	approximation, there exists a  Dirac   node at the    $\Gamma$
	($\vec{k}=0$) point centered at zero  energy. On including the
	effect of the curvature terms,  $\Delta H$,  we have seen that 
	a gap is  induced
	and the spectrum near  zero energy is  that of a massive Dirac
	particle,  that  retains  particle-hole  symmetry,  and  whose
	dispersion  is centered  near   the  $\Gamma$ point. All   the
	negative energy  states are then occupied  (in the `d' particle
	representation of quasiparticles that has been adopted) since  
	the quasiparticle chemical  potential, in the absence of Zeeman 
	splitting, is at zero energy.

	This   situation is topologically  identical  to  a free Dirac
	particle in two dimensions, with a  mass term. In other words,
	the  spectra of the two systems   can be continuously deformed
	into one another {\it without} closing the gap at the chemical
	potential.  The  massive Dirac equation  in two dimensions has
	the dispersion $E(p)=\pm \sqrt{p^2   + m^2}$ and the  negative
	energy branch is completely filled in the ground state. There,
	it  is well known   that the  zero temperature Hall   conductance of the  Dirac
	particles    is          quantized        $\sigma_{xy}=\frac12
	\frac{e^2_D}{h}\sgn{m}$  where  $e_D$ is the charge associated
	with      the     Dirac         particle        \cite{Haldane}
	\cite{mpaf}.  Superconductor  quasiparticles of  course do not
	carry a well defined electrical charge - however, for the case
	of interest the component of  the quasiparticle spin along the
	applied  magnetic field is  conserved,  and plays  the role of
	$e_D$; as  we  have seen  the  density of  the conserved   `d'
	particles corresponds to the spin density  in the direction of
	the  field. Hence, in the  present case  of quasiparticles we expect a
	quantized    spin     hall     conductance    \cite{Volovik}
	\cite{Senthil} with $e_D$    replaced by the quantum  of  spin
	$\hbar/2$ carried by the quasiparticles. Thus, if we define 
	$\sigma_0^{\mbox{s}} = \frac{(\hbar/2)^2}h$, 

\begin{eqnarray}
\sigma_{xy}^{\mbox{s}} &=& \frac12 \sigma_0^{\mbox{s}}\sgn{m_D}\\
\sigma_{xx}^{\mbox{s}} &=& 0
\end{eqnarray}

for a single node. 

	It  is  easily seen  that  for  inversion  symmetric lattices,
	opposite      nodes  contibute equally to       the spin  hall
	conductivity.   The     Hamiltonian   at   node     $\bar{1}$,
	$H_{\bar{1}}+\Delta    H$,   under   the   inversion operation
	$\psi_{\bar{1}}(\vec{r})                           \rightarrow
	\psi_{\bar{1}}(-\vec{r})$,   is  found    to be  identical  to
	$H_{1}+\Delta   H$.   Therefore     they have    the same  gap
	$|m_D(1)|=|m_D(\bar{1})|$, and since  the Hall  conductance is
	unchanged by  the inversion (rotation by  an  angle of $\pi$),
	have         identical    spin     Hall           conductances
	$\sigma_{xy}^{\mbox{s}}(\bar{1})=\sigma_{xy}^{\mbox{s}}(1)$. Similarly
	we can relate the  gaps   and spin Hall conductances   between
	nodes $2$ and $\bar{2}$ \cite{antinodemD}.

	If no  further symmetry of  the vortex lattice is assumed, the
	size  of the    gap  and the    sign of  the  contribution  to
	$\sigma_{xy}^{\mbox{s}}$  from  the   nodes  $2$  and
	$\bar{2}$ are not simply related  to their values at the  nodes
	$1$  and $\bar{1}$. Thus,  as regards the low temperature Hall
	transport,  two   scenarios present themselves.  The  first in
	which the contribution from both pairs of  nodes have the same
	sign,  in  which    case  $\sigma_{xy,Tot}^{\mbox{s}}     =
	4\sigma_{xy}^{\mbox{s}}(1)               =                2
	\sigma_0^{\mbox{s}}\sgn{m_D(1)} $   and   the  second, when
	contributions  from the two pairs   of nodes are opposite  and
	cancel,                 $\sigma_{xy,Tot}^{\mbox{s}}       =
	2\sigma_{xy}^{\mbox{s}}(1)+2\sigma_{xy}^{\mbox{s}}(2)=0$,
	and  the spin Hall  conductance is quantized  to zero. In both
	cases the quantization begins to set  in at temperatures lower
	than the   smallest  gap.  These two   cases are topologically
	equivalent  to    homogenous  superconductors   with   pairing
	wavefunctions   that    break      time   reversal   symmetry,
	d$_{x^2-y^2}$+id$_{xy}$ in the  first case and d$_{x^2-y^2}$+i$s$, or any other thermal insulator, in
	the second case.

	When the vortex lattice posesses  an additional symmetry, that
	of reflection about an axis  that bisects the nodal directions
	(dashed lines in  Figure \ref{nodes}), we can relate the physics at the nodes ($1$ $\bar{1}$) with that at the nodes ($2$ $\bar{2}$). We then have $|m_D(1)|=|m_D(2)|$
	and the spin hall  conductances from all  the nodes add. Thus,
	the  first  scenario, with  $\sigma_{xy,Tot}^{\mbox{s}} = 2
	\sigma_0^{\mbox{s}}\sgn{m_D(1)}   $  will be realized.  For
	lattices that weakly  break this symmetry, the exact  equality
	of the gaps $|m_D(1)|$ and $|m_D(2)|$ no longer holds, but the
	spin   hall conductance is still  expected to  remain   equal to $ 2
	\sigma_0^{\mbox{s}}\sgn{m_D(1)} $. A finite deformation
	of  the lattice is at least required to make a transition to 
	the topologically distinct state with $\sigma_{xy,Tot}^{\mbox{s}}=0$.

{\it Effect  of  Zeeman Splitting  on  Quantization:}  So  far we  have
neglected  the Zeeman splitting of  the quasiparticles by the magnetic
field, which, as we  have seen earlier, plays the  role of the chemical
potential  for   the  `d' particles.  The   magnitude  of   this term,
$E_{Zeeman}=  \frac12 g\mu_B B  =$0.7 KTesla$^{-1}$ is of the
same order     as   the  expected    mass  gap   in  the vortex state of YBCO $m_D  \sim
E_1^2/E_F=$0.5KTesla$^{-1}$   and     has      the       same    field
dependence. Therefore, the Zeeman   term  plays an important  role  in
determining whether  or not the  exact quantization  of  the spin Hall
conductance discussed  above will be  realised in these materials. If
the quasiparticle   gap  ($|m_D|=\min[|m_D(1)|,|m_D(2)|]$) exceeds the
Zeeman term $|m_D|>  |E_{Zeeman}|$ then, the quasiparticle chemical potential lies in the gap, and quantization  of the
spin  hall conductance is   expected,  although the temperature  below
which this will set in is now given by the difference $|m_D|-|E_{Zeeman}|$. If,
however,  the Zeeman splittling  exceeds   the gap, then the  chemical
potential  for  the quasiparticles   lies in a  region where  extended
states  are present, and the   exact quantization will be lost.  (When
disorder  is taken into account, these states may   be  localised and then 
exact quantization  will be recovered).  The question of whether the
quasiparticle gap exceeds the Zeeman energy depends on several details
of the problem such  as the value  of the anisotropy $\alpha$ and  the
vortex lattice geometry, as well as on the value  of the band mass and
$g$ factor of the electron for the material in consideration.

{\it  Quasiparticle Heat   Transport:}  While   so  far   we have  been
discussing   the spin conductance of  the   quasiparticles, a far more
accessible transport  quantity in experiments is  the thermal   conductance.  In the
superconductor, heat  is transported by the  quasiparticles as  well as
the phonons. The electronic part of the thermal conductivity is usually isolated in one of the following ways. For the longitudinal part of the thermal conductivity ($\kappa_{xx}$), the phononic contribution at low temperatures varies as $T^3$\cite{Taillefer} and can typically be separated out. As for the   thermal Hall conductance, that is expected to arise
purely from  the quasiparticles, since phonons are  not expected to be
skew scattered by the magnetic field \cite{Ong0}.

	At  low temperatures, a   Widermann-Franz relation between the
	themal and spin conductances of the quasiparticles is expected
	to hold:
\be
\mbox{Lim}_{T\rightarrow 0}\frac{\kappa}T = (\frac{\pi^2}3\frac{k_B^2}{(\hbar/2)^2})\sigma^{\mbox{s}}
\ee

Thus, for the cases  discussed earlier we are  led  to expect, at  low
temperatures,  a `quantized'  thermal  Hall conductivity, in the  sense
$\kappa_{xy}/T  \rightarrow  n  (\frac{\pi^2}3\frac{k_B^2}{h})$ where
n=$\pm   2$,$0$ for  the   two  possible  cases,  and   $(\kappa_{xx}/T)
\rightarrow 0$.

{\it Violation of Simon-Lee Scaling} In \cite{Simon&Lee}, Simon and Lee
showed that the leading contribution to  the thermal Hall conductivity
takes the scaling form: 
\be
\label{SL}
\kappa_{xy} \sim T^2 F_{xy}(T/H^\frac12 \sqrt{v_F v_\Delta})
\ee

Clearly, the quantized   Hall conductivity   derived above does    not
satisfy this scaling relation. The reason for this is easily seen. In
deriving  equation (\ref{SL})  for   the vortex lattice  following the
arguments in  \cite{Simon&Lee}, we need  to assume that the spectrum of
states in the linearized theory at a given crystal momentum, are well
separated in energy,  and so are only weakly  mixed by  the perturbing
curvature terms ($\Delta H$). While this assumption  is true over most
of the Brillouin zone,  it breaks  down at  the isolated points  where
doubly degenerate  states are present. It is  precisely the lifting of
this degeneracy  by $\Delta H$  that  leads to  the mass  term for the
Dirac quasiparticles, and the quantized thermal  Hall conductivity that violates the scaling form above. For
temperatures exceeding the induced gap $m_D \sim E_1^2/E_F$, Simon-Lee
scaling is recovered.

\subsection{Numerical Evaluation of Gaps:}
	In order to make the preceeding discussions more concrete, we present here a numerical evaluation of the gaps induced by the curvature terms. These results confirm our assumption that their effect on the spectrum of the linearized Hamiltonian is indeed small. We consider the simple situation of a square lattice of $\frac{hc}{e}$ (double) vortices, oriented along the nodal directions. The mass term induced by $\Delta H$ for the Dirac cone at zero energy is calculated (at node 1) for two values of the anisotropy parameter ($\alpha$=1,2). Reflection symmetry of the lattice about the 45$^0$ line allows us to relate the mass term at the other nodes to $m_D(1)$. The linearised Hamiltonian to consider is:
\be
H_1 = v_F(p_x \sigmaz + \alpher p_y \sigmax + \hat{x}\cdot \vec{P}_s(\vec{r})\One)
\ee 

where the explicit form of $\vec{P}_s(\vec{r})$ that we use is:
\be
\vec{P}_s(\vec{r}) = \frac{h}{d^2} \sum_{\vec{G}\not= 0} \frac{\hat{z}\times\vec{G}}{|\vec{G}|^2}\mbox{e}^{i\vec{G}\cdot\vec{r}}
\ee

where $\vec{G}=\frac{2\pi}{d}$(m, n) are the reciprocal vectors for a square vortex lattice with lattice parameter $d$, which from flux quantization for a $\frac{hc}e$ vortex lattice satisfies $d=\sqrt{h/eB}$. This form for $\vec{P}_s$ may be derived from equation (\ref{curlps2}) if we assume that (a) $\vec{P}_s$ is divergence free i.e. $\nabla \cdot \vec{P}_s(\vec{r})=0$, which is valid if the superfluid density can be taken as uniform and (b) the magnetic field is constant, which is justified if the penetration depth $\lambda$ is large compared to the magnetic length $d/\lambda\ll 1$.

\begin{figure}
\epsfxsize=3.2in
\centerline{\epsffile{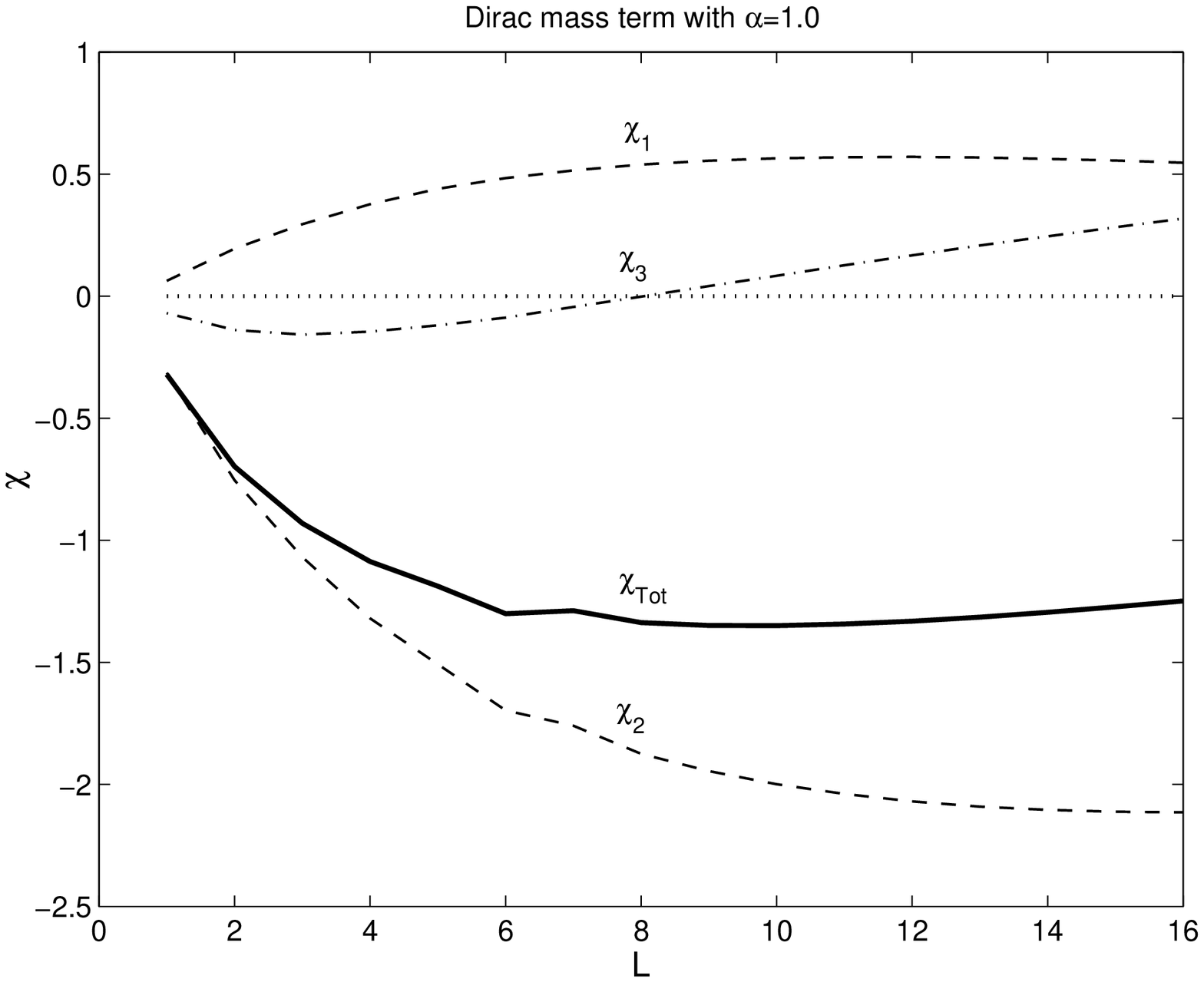}}
\epsfxsize=3.2in
\centerline{\epsffile{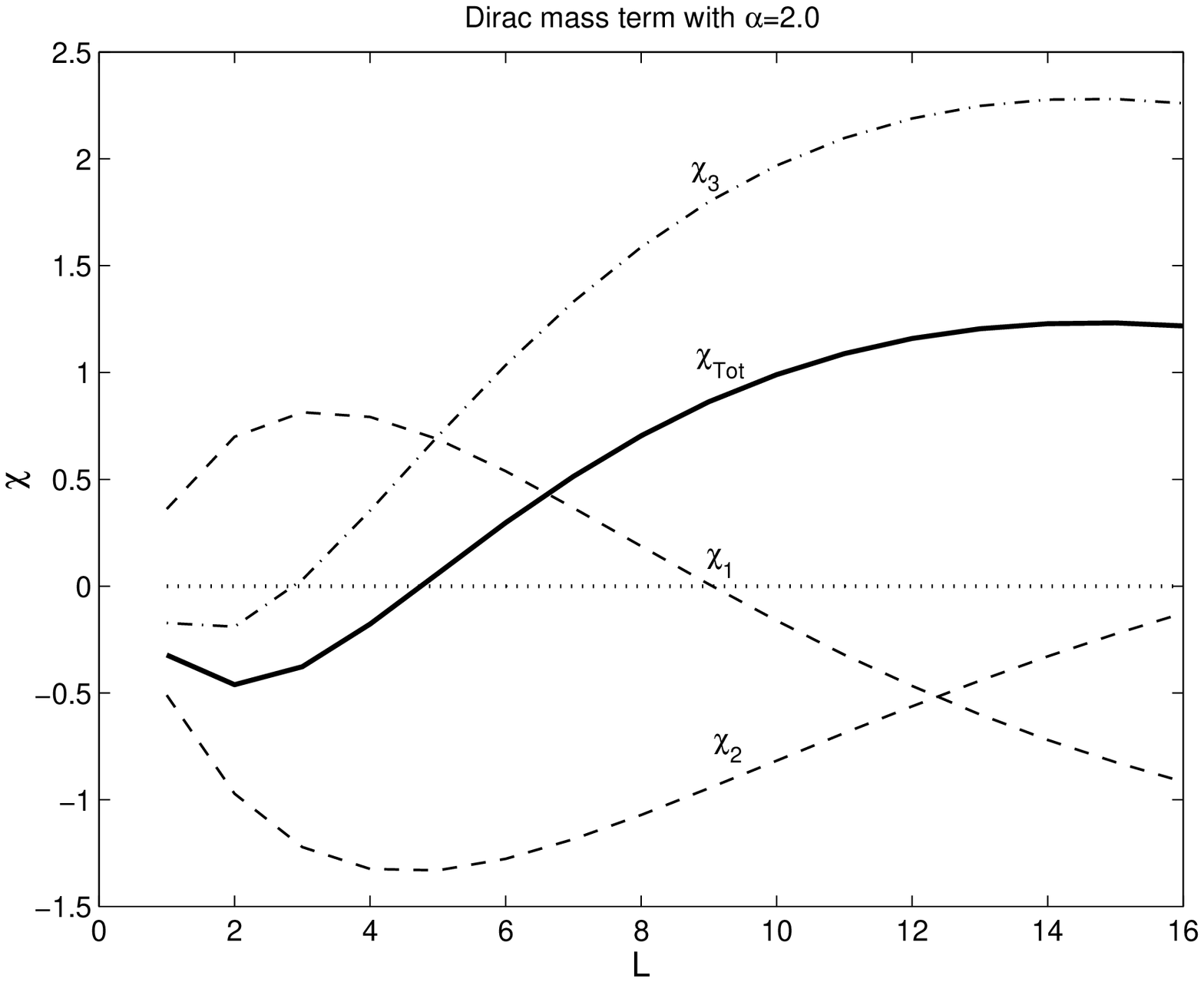}}
\caption{Numerical evaluation of the Dirac mass term for a square vortex lattice of $hc/e$ vortices for two values of the anisotropy $\alpha=1.0$, $2.0$. The magnitude of the dimensionless quantity $\chi_{Tot}$ controls the gap size, while its sign determines the sign of the quantized thermal Hall effect. Here, $\chi_{Tot}$, as well as the separate contributions $\chi_i$ arising from the different curvature terms are plotted as a function of $L$, that is a measure of the number of reciprocal lattice vectors $(2L+1)^2$, retained in the numerical calculation.}
\label{chi_vs_L}
\end{figure}

By our previous arguments we are assured of a pair of degenerate states at $E=0$ for this linearized Hamiltonian, i.e. there exists a pair of linearly independent wavefunctions that satisfies $H_1\psi(\vec{r})=0$. The pair of degenerate states are obtained numerically, and the effect of $\Delta H$ is evaluated in degenerate perturbation theory within this two dimensional subspace, to yield the Dirac mass $m_D$ (\ref{m_D}). Thus, the curvature terms: 
\be
\Delta H = \frac1{2m} p^2 \sigmaz + \frac1{2m}\vec{P}_s^2(r)\sigmaz + \frac1m \vec{p}\cdot\vec{P_s}\mbox{\boldmath $1$}+\{\frac{\Delta_0}{p_F^2}p_x p_y \sigmax\}
\ee
generates a Dirac mass at this node
$$
m_D = m_1 + m_2 + m_3 + \{ m_4 \} 
$$
where each term $m_i$ results from the corresponding piece in $\Delta H$. The term $m_4$, which is smaller by a factor of $\Delta/E_F$, can be neglected. The remaining terms can be written in the following form that explicitly displays their dependence on the magnetic field (B) and band mass (m). 
\begin{eqnarray*}
m_{1,2,3} &=& \chi_{1,2,3} \frac{\hbar eB}{2m}\\
\chi_{Tot}&=& \chi_1+\chi_2+\chi_3\\
m_D &=& \chi_{Tot} \frac{\hbar eB}{2m}
\end{eqnarray*}
where the $\chi_i$ are dimensionless quantities that depend only on the lattice geometry and the value of the anisotropy $\alpha$. Therefore it is convenient to present the numerical results in terms of them. Physical quantities of interest are easily related to $\chi_{Tot}$. Its magnitude sets the size of the gap while its sign determines the sense of the quantized thermal Hall effect. Thus, the gap is given by $2|m_D|= 2|\chi_{Tot}\frac{\hbar eB}{2m}|$ and is the same at all nodes due to the reflection symmetry about the 45$^0$ line of the square lattice being considered. The thermal Hall response from all these nodes add to give the quantized thermal Hall conductance:
$$
\kappa_{xy}/T = 2 \{ \frac{\pi^2k_B^2}{3 h}\mbox{sgn}(\chi_{Tot}\frac{eB}{m})\}
$$ 
We formulate the numerical problem in reciprocal (momentum) space, retaining (2L+1)$^2$ reciprocal lattice vectors. The null space of $H_1$ ($H_1 \psi=0$) at the $\Gamma$ point is obtained and the Dirac mass terms calculated using equations (\ref{ndeltaH},\ref{nk},\ref{m_D}). These are shown in Figure \ref{chi_vs_L} for two values of $\alpha$, ($\alpha$=1,2), where $\chi_{Tot}$ as well as the separate contribution of the different curvature terms, the $\chi_i$s are plotted as a function of L. We see that satisfactory convergence is obtained by L=16. The magnitude of the Dirac mass, in both cases turns out to be roughly $|m_D|\sim  1.2 |\frac{e\hbar B}{2m}|$ (which is 0.8KTesla$^{-1}$, if $m$ is taken as the electron mass). Thus, for fields of a few Tesla, the gap is much smaller than the energy scale $E_1$ (50KTesla$^\frac12$ in YBCO), which justifies our treatment of the curvature terms as perturbations on the linearized Hamiltonian. Incidentally, the mass term in this particular example, exceeds the Zeeman energy (0.7KTesla$^{-1}$ with g=2), and so the chemical potential for the quasiparticles will lie within the gap.

	The overall sign of the quantized thermal Hall conductivity is clearly seen to be a function of $\alpha$, it reverses sign on going from $\alpha=1$ to $\alpha=2$ ($\chi_{Tot}$ takes opposite signs for these two values of $\alpha$). Its dependence on the sign of $\frac{eB}{m}$ is expected, since on inverting the sign of any of these ($e$, $B$, $m$), the quasiparticles are expected to deflect in the opposite direction.

	Thus, given a real material whose vortex lattice structure and microscopic parameters are known, one can follow the recipe above to (a) calculate if a quantized Hall conductance is expected for the clean lattice case (i.e. does $m_D$ exceed $E_{Zeeman}$), and if so (b) the quantization value ($\pm$2,0 in appropriate units), and the temperature scale (set by $|m_D|-|E_{Zeeman}|$) below which quantization is expected.

	We have already noted the similarity of the state with quantized thermal Hall conductivity of $\pm$2 to the pure d$_{x^2-y^2}$ $\pm$ id$_{xy}$ system - both are gapped and have identical low temperature thermal conductivities. Other authors, \cite{Laughlin}\cite{TVR} have considered the possibility of a magnetic field inducing a `id$_{xy}$' component in a d$_{x^2-y^2}$ superconductor.
We believe that our work for the vortex lattice provides a concrete realization of the general ideas of \cite{Laughlin}\cite{TVR}. Further, we have described a proceedure that allows us to determine {\it when} such a state is to be expected, and the size of the energy gap induced.

\section{Discussion}
In summary, we have considered the problem of d-wave quasiparticles in the mixed state within the perfect lattice approximation. Within the linearized theory  we find a massless Dirac dispersion for the low energy quasiparticles, so long as the vortex lattice posesses inversion symmetry. On going beyond the linearized approximation and including the smaller curvature terms, a small gap is found to open and the low energy excitations now behave like massive Dirac particles. The size of the gap is proportional to the applied magnetic field, and is roughly of order $E_1^2/E_F$ ($\approx 0.5$ KTesla$^{-1}$ in YBCO). If the sign of the Dirac mass term (appropriately defined) is identical at the four nodal points, then we obtain a topologically non-trivial state - i.e. one that has gapless chiral quasiparticle modes at the edge, which in this case are the same as in a pure d$_{x^2-y^2}\pm i$d$_{xy}$ superconductor. When the chemical potential for quasiparticles lies within the gap a quantized thermal Hall conductivity of $\kappa_{xy}/T = \pm 2 \frac{\pi^2k_B^2}{3h}$ is expected at low temperatures. The chemical potential for these quasiparticles is controlled by the Zeeman energy $E_Z$, and if this is smaller than the gap scale $m_D$ the quantized thermal Hall effect will be observed at low temperatures ($T\ll [m_D - E_Z]$). We also described a computationally simple proceedure for evaluating the numerical value of the gap and the quantized themal Hall conductance, which depend on the microscopic material parameters as well as the value of the anisotropy and the geometry of the vortex lattice.   

Although in general the quantized thermal Hall conductivity can take on any integer value (see Appendix D), on the basis of the energetics for this particular situation  we have argued that the thermal Hall coefficient is quantized to one of the discrete set of values ($\kappa_{xy}/T=\pm 2$, $0$ in appropriate units). This is a consequence of (a) the independent node approximation,  and (b) that at each of the four nodes, the only low energy feature (compared to $E_1^2/E_F$) is a single Dirac cone centered at zero energy within the linearized problem for inversion symmetric lattices. For special situations where the linearized theory may posess additional low energy features, for example as reported in \cite{FT,Marinelli,Kallin} for the square lattice of vortices at large anisotropy ($\alpha$) - other integer values of the quantized thermal Hall conductance may be realized.

	Throughout, we have treated the four nodes as being independent, and neglected the effect of inter-node scattering. This is expected to be important when there is a high degree of commensuration between the reciprocal vectors associated with the vortex lattice, and the momenta separating the four nodes at the Fermi surface. In that case, as demonstrated in \cite{Vafek} the nodal points acquire a gap due to the formation of a quasiparticle density wave. However, away from such special commensuration, the independent node approximation used here is expected to hold.
	
	Our discussion in this paper has been restricted to the clean limit but of course disorder is always present in the physical system. The topological character of the states we have been considering renders them insensitive to the effects of weak disorder. Further, if the quasiparticle chemical potential lies outside the gap, while a quantized thermal Hall effect would not be expected in the pure case, disorder could localises the quasiparticle states at the chemical potential stabilizing the quantization. However, for larger disorder strengths, a completely different approach that does not rely on the Bloch nature of the quasiparticle states will be required. With these caveats in mind we turn to survey some of the relevant experiments in YBCO.  

	Low temperature thermal-Hall measurements would provide the most direct test of the occurance of such topologically ordered states in superconductors. Currently, such measurements on YBa$_2$Cu$_3$O$_7$ go down to temperatures of about 12K in a field of 14Tesla \cite{Ong}, which is presumably still too high to observe the quantization, should it exist. Indeed, although the measured $\kappa_{xy}$ at a given temperature is found to saturate for stronger fields, the value at this plateau scales as $T^2$, rather than linear in $T$ as would be expected if $\kappa_{xy}/T$ was quantized. However it is an intriguing fact that the plateau value of $\kappa_{xy}$ at the lowest temperature measured (12.5K) is very  close to what would be expected from a quantized thermal Hall conductance of $|\kappa_{xy}/T|=2$ (in appropriate units). While this experiment is not conclusive with regard to the low temperature state of the quasiparticles, we can look to specific heat measurements for a signature of the chemical potential lying in a gap. Such low temperature specific heat measurements down to 1K in magnetic fields of 14Tesla in YBa$_2$Cu$_3$O$_7$ have been reported in\cite{Junod}. However, they show no evidence of a gap, in fact a finite density of quasiparticle states that scales as the square root of the field was found. In principle, a quantized thermal Hall condutance could still arise in such a system, if the states at the chemical potential are localized - in all cases the quantization of the thermal Hall conductance will be accompanied by the vanishing of the longitudinal thermal conductance $\kappa_{xx}/T \rightarrow 0$ at temperatures below the (mobility) gap scale. But low temperature $T<0.1$K longitudinal thermal conductivity measurements in fields upto 8Tesla in YBa$_2$Cu$_3$O$_{6.9}$ \cite{Taillefer} reveals $\kappa_{xx}/T$ saturating to a nonzero value that rules out a quantized thermal Hall effect in this material down to these low temperatures. Thus, for this case of YBCO, either the Zeeman splitting causes the chemical potential to lie outside the gap region, or else the vortex lattice in this case is so disordered that the perfect lattice assumption we start with requires serious modification. Nevertheless, given the large number of potentially different experimental systems with a d-wave gap that are available, it is not unreasonable to expect that the quantized thermal Hall effect, realized in the manner described, will be observed in the future. 
\section*{Acknowledgements} The author thanks F.D.M. Haldane for several stimulating conversations and L. Balents, B. Halperin, D. Huse, M. Krogh, L. Marinelli, A. Melikyan, N.P. Ong, T. Senthil, Z. Tesanovic and O. Vafek for useful discussions. Support from NSF grant DMR-9809483 and a C.E. Procter fellowship, as well as hospitality at the ITP, Santa Barbara where part of this work was done, are acknowledged.    

\section*{Appendix A: Perturbation Theory}
As pointed out in \cite{Marinelli}, perturbation theory  may be successfully employed in proving the existence of zero energy states of the linearized Hamiltonian. Here we consider two cases within this technique. First, we look at general inversion symmetric vortex lattices of $hc/e$ (double) vortices, where our expectation of a degenerate pair of zero energy states is confirmed within this technique. Second, we consider a specific example of a vortex lattice of $hc/2e$ vortices, which is invoked later to illustrate some of the subtle features of the Franz-Tesanovic transformation when used within the linearized approximation.
 
{\em Zero Energy States for Vortex Lattice of $hc/e$ Vortices:}
	Here, we consider an inversion symmetric vortex lattice of $hc/e$ (double) vortices and establish the existence of a pair of zero energy states to all orders in perturbation theory.
	Essentially this is the problem of a Dirac particle in the presence of a periodic scalar potential $V(x,y) \equiv P_{sx}(x,y)$:
$$
V(\vec{r}+\vec{e}_i) = V(\vec{r})
$$
where $\vec{e}_i \in \{\vec{e}_1, \vec{e}_2 \}$ are the translations that generates the vortex lattice and 
\be
\label{odd}
V(-x,-y)= - V(x,y)
\ee
which results from the inversion symmetry of the vortex lattice. The Hamiltonian that we consider is
\be
H = v_F \{\slash{\mbox{\boldmath $p$}} + V(x,y)\}
\ee
where, by a slight abuse of notation we have found it convenient to define:
\begin{eqnarray*}
\slash{\mbox{\boldmath $p$}} & \equiv & p_x \sigmaz + \alpher p_y \sigmax \\
				&=& \sigmaz(-i\partial_x) + \alpher \sigmax(-i\partial_y) 
\end{eqnarray*}

To find the zero energy states we solve the equation 
$$
H \psi(x,y)=0
$$
at the $\Gamma$ point, i.e. with periodic boundary conditions for the wavefunction. We choose to work in reciprocal space, where, as usual, the reciprocal vectors $\vec{G}_{\vec{n}} = n_1\vec{g}_1 + n_2\vec{g}_2$, with integers $\vec{n}=(n_1,n_2)$, and basis vectors ($\vec{g}_i$) that satisfy the conditions:
$$
\vec{g}_i \cdot \vec{e}_j = 2\pi \delta_{ij}
$$
The potential and the wave function may be Fourier expanded as:

\begin{eqnarray*}
\psi(\vec{r}) &=& \sum_{\vec{n}} \tpsi (\vec{G}_{\vec{n}}) \mbox{e}^{i\vec{G}_{\vec{n}}\cdot \vec{r}}\\  
V(\vec{r}) &=& \sum_{\vec{n}} \tV (\vec{G}_{\vec{n}}) \mbox{e}^{i\vec{G}_{\vec{n}}\cdot \vec{r}}
\end{eqnarray*} 

and in this basis the eigenvalue equation for the zero energy states is:
\be
\label{zeromode}
\slash{\mbox{\boldmath $G$}}_{\vec{n}} \tpsi(\vec{G}_{\vec{n}}) + \sum_{\vec{n}'} \tV(\vec{G}_{\vec{n}}-\vec{G}_{\vec{n}'})\tpsi(\vec{G}_{\vec{n}'})=0
\ee
 From the fact that the potential $V(x,y)$ is odd under inversion (\ref{odd}), we have that the Fourier components $\tV$ are also odd functions of the reciprocal lattice vectors:
\be
\label{oddG}
\tV(-\vec{G})=-\tV(\vec{G})
\ee
and in particular, the zero wave-vector component vanishes $\tV(0)=0$.

We now prove to all orders in perturbation theory that there exists a pair of states that satisfies the above equation (\ref{zeromode}). We treat $V$ as a perturbation on the free Dirac particle, and to organize the perturbation expansion it is convenient to imagine that there is a small control parameter $\epsilon$, $V \rightarrow \epsilon V$ which is set equal to one at the end of the calculation. Then, if we can expand the solution in powers of $\epsilon$ we obtain:
\be
\tpsi (\vec{G}) = \tpsi_0(\vec{G}) + \epsilon \tpsi_1(\vec{G}) + \epsilon^2 \tpsi_2(\vec{G})+...
\ee
. Inserting this in equation (\ref{zeromode}) we can solve to each order in $\epsilon$  to obtain, for $\vec{G} \not= 0$, the following recursion relation:
\be
\label{recursion}
\slash{\mbox{\boldmath $G$}} \tpsi_M(\vec{G}) = - \sum_{\vec{G}'}\tV(\vec{G}-\vec{G}')\tpsi_{M-1}(\vec{G}')
\ee
where we have used a more compact notation replacing $\vec{G}_{\vec{n}}, \vec{G}_{\vec{n}'}$ by $\vec{G}, \vec{G}'$. The equation for $\vec{G} = 0$ imposes on the solution the condition:
\be
\label{condition}
0 = \sum_{\vec{G}'}\tV(-\vec{G}')\tpsi_{M-1}(\vec{G}')
\ee

We start by finding $\psi_0$, i.e. solving the free problem, and then use the recursion relation (\ref{recursion}) to iteratively obtain $\psi_M$. A zero energy solution exisits if the condition (\ref{condition}) is satisfied at all orders of perturbation theory, i.e. if (\ref{condition}) is satisfied by the $\tpsi_M$ so obtained, for all $M$. The solution to the free problem is trivial, there are two zero energy states corresponding to the spatially constant function times any spinor, namely $\tpsi_0(\vec{G})=0$ if $\vec{G} \not=$ 0 and $\tpsi_0(0)$= (1 0)$^T$ or (0 1)$^T$. To this order of perturbation theory, the condition (\ref{condition}) is satisfied, since $\tV(0)=0$. Now, examinining the recursion relation (\ref{recursion}) and using (\ref{oddG}), it is easily seen that at any order in perturbation theory the wavefunction is an {\it even} function of $\vec{G}$ - that is
\be
\tpsi_M(\vec{G}) = \tpsi_M(-\vec{G})
\ee
 and hence the condition (\ref{condition}) for the existence of a zero energy state is automatically satisfied at all orders of perturbation theory:
\begin{eqnarray*}
\sum_{\vec{G}'}\tV(-\vec{G}')\tpsi_{M-1}(\vec{G}') &=& \frac12 \sum_{\vec{G}'}[\tV(-\vec{G}')\tpsi_{M-1}(+\vec{G}') \\
	& & + \tV(+\vec{G}')\tpsi_{M-1}(-\vec{G}')]\\
	  &=& 0
\end{eqnarray*}
In fact, the two choices of $\tpsi_0$ give rise to two distinct zero energy states ($\phi$ and $\psi$) that are related by $\tilde{\phi}(\vec{G})=\TAU \tpsi^*(-\vec{G})$, which are easily seen to be orthogonal. 

	Thus, we have been able to confirm to all orders in perturbation theory that the linearized Hamiltonian for an inversion symmetric vortex lattice of $hc/e$ vortices posesses a pair of degenerate states at zero energy.

{\it Zero Energy States of a $hc/2e$ Vortex Lattice:} We now consider a vortex lattice of $hc/2e$ vortices, specifically, we look at a vortex lattice with two vortices per unit cell but consider making a Franz-Tesanovic transformation that artificially doubles the unit cell, as shown in Figure \ref{fig4}b. The reason for considering this particular case is that it will help us expose some subtle features of the Franz-Tesanovic transformation when used in combination with the linearized approximation, in the absence of a proper regularization. While we shall find a pair of zero energy states for this case, numerical work on an essentially equivalent situation \cite{Marinelli} - the same lattice, only treated using a Franz-Tesanovic gauge transformation that preserved the two vortex unit cell (like in figure \ref{fig4}a) - found a gap at zero energy. This paradoxical situation is taken up in more detail in the following two appendices where a spurious coupling present in the unregularised (continuum) theory is found to be responsible for this state of affairs.

The linearized problem is that of a Dirac particle in the presence of a periodic scalar potential $V(x,y) \equiv P_{sx}(x,y)$ and a vector potential $\vec{a}$ (chosen to be periodic) corresponding to solenoids of flux $\pm \pi q$ located at the positions of the vortices $\vec{r}_{\pm \pm} = \pm\frac{d}2 \pm \frac{\vec{\rho}}2$ as shown in Figure \ref{fig4}. Thus, the linearized Hamiltonian we consider is:
\be
H = v_F\{(\slash{\mbox{\boldmath $p$}}+\slash{\mbox{\boldmath $a$}}) + V(x,y)\}
\ee
where the vector potential $\vec{a}$ satisfies:
\begin{eqnarray}
\vec{\nabla}\times\vec{a}(\vec{r}) &=& \pi\hbar\sum_n[\delta(\vec{r}-\vec{R}_n -\vec{r}_{++}) - \delta(\vec{r}-\vec{R}_n -\vec{r}_{+-}) \nonumber \\
	& & - \delta(\vec{r}-\vec{R}_n - \vec{r}_{-+}) + \delta(\vec{r}-\vec{R}_n +\vec{r}_{--})]
\end{eqnarray}
due to the inversion symmetry of this vortex lattice we have:
$$
V(-\vec{r})=-V(\vec{r})
$$
and can choose the vector potential so that
$$
\vec{a}(-\vec{r}) = -\vec{a}(\vec{r})
$$

Also, as a consequence of the artificial doubling of the unit cell we note the relations:
\begin{eqnarray}
V(\vec{r}+\hat{x}d) &=& V(\vec{r})\\
\vec{a}(\vec{r}+\hat{x}d) &=& -\vec{a}(\vec{r})
\end{eqnarray}

To find the zero energy states we have to solve the equation 
$$
H \psi(x,y)=0
$$
at the $\Gamma$ point, i.e. with periodic boundary conditions for the wavefunction. Denoting the Fourier components of the various quantities by a tilde as before, we have the equivalent equation in reciprocal space:

\be
\label{zeromode1}
\slash{\mbox{\boldmath $G$}} \tpsi(\vec{G}) + \sum_{\vec{G}'} [\slash{\mbox{\boldmath $\ta$}}(\vec{G}-\vec{G}') + \tV(\vec{G}-\vec{G}')]\tpsi(\vec{G}')=0
\ee

The fact that the potentials are odd functions of $\vec{r}$ results in:
\begin{eqnarray}
\label{oddG1}
\tV(-\vec{G}) &=& -\tV(\vec{G})\\
\ta(-\vec{G}) &=& -\ta(\vec{G})
\end{eqnarray}

We consider both the vector and scalar potentials as perturbations of order $\epsilon$; $V \rightarrow \epsilon V$ and $\vec{a} \rightarrow \epsilon \vec{a}$, and expand the wavefunction in powers of $\epsilon$:
$$
\tpsi(\vec{G}) = \sum_{M=0}^{\infty}\epsilon^M\tpsi_M(\vec{G})
$$
. Inserting this in equation (\ref{zeromode}) we can solve to each order in $\epsilon$ which obtains for us (for $\vec{G} \not= 0$) the following recursion relation:
\be
\label{recursion1}
\slash{\mbox{\boldmath $G$}} \tpsi_M(\vec{G}) = - \sum_{\vec{G}'} [\slash{\mbox{\boldmath $\ta$}}(\vec{G}-\vec{G}') + \tV(\vec{G}-\vec{G}')]\tpsi_{M-1}(\vec{G}')
\ee
 while the equation for $\vec{G} = 0$ imposes on the solution the condition:
\be
\label{condition1}
0 = \sum_{\vec{G}'} [\slash{\mbox{\boldmath $\ta$}}(\vec{G}-\vec{G}') + \tV(\vec{G}-\vec{G}')]\tpsi_{M-1}(\vec{G}')
\ee

Once again, we start the perturbation theory by finding $\psi_0$, i.e. solving the free problem, and then use the recursion relation (\ref{recursion1}) to iteratively obtain $\psi_M$. A zero energy solution exisits if the condition (\ref{condition1}) is satisfied at all orders of perturbation theory. The solution to the free problem is trivial, there are two zero energy states corresponding to the spatially constant function times any spinor, namely $\tpsi_0(\vec{G})=0$ if $\vec{G} \not=$ 0 and $\tpsi_0(0)$= (1 0)$^T$ or (0 1)$^T$. To this order of perturbation theory, the condition (\ref{condition1}) is satisfied, since by equation (\ref{oddG1}), $\tV(0)=0$ and $\ta(0)=0$. Now, examining the recursion relation (\ref{recursion1}) and using (\ref{oddG1}), it is easily seen that at any order in perturbation theory the wavefunction is an {\it even} function of $\vec{G}$ - that is
\be
\tpsi_M(\vec{G}) = \tpsi_M(-\vec{G})
\ee
 and hence the condition (\ref{condition1}) for the existence of a zero energy state is automatically satisfied at all orders of perturbation theory:
$$
\sum_{\vec{G}'}[\slash{\mbox{\boldmath $\ta$}}(-\vec{G}')+\tV(-\vec{G}')]\tpsi_{M-1}(\vec{G}')=0 
$$

In fact, the two choices of $\tpsi_0$ give rise to two distinct zero energy states ($\phi$ and $\psi$) that are related by $\phi(\vec{r})=\TAU \psi^*(\vec{r}+\hat{x}d)$. That these wavefunctions are linearly independent is shown in the next appendix. 

Thus, for this situation involving $hc/2e$ vortices,we have shown to all orders of perturbation theory that the (continuum) linearized Hamiltonian in the particular Franz-Tesanovic transformation used above, has a degenerate pair of zero energy states. 

\section*{Appendix B: Regularizing the Linearized Theory}
In this appendix we consider some subtle features of the Franz-Tesanovic transformation that arise in the linearized approximation. Recall, that the ficticious U(1) gauge field $\vec{a}$ was introduced only to take into account the statistical interaction of the quasiparticles and $hc/2e$ vortices. This led to the Dirac Hamiltonain:
\be
\label{hamlt}
H^{FT}_1 = v_F\{(p_x + a_x)\sigmaz + \alpher (p_y+a_y)\sigmax+\hat{x}\cdot\vec{P}_s(\vec{r})\}
\ee

with
$$
\vec{\nabla}\times \vec{a}(\vec{r}) = \hbar\pi \sum_i q_i \delta^{(2)}(\vec{r}-\vec{r}_i)
$$
where $q_i$ is an odd integer. Now, as is well known from the theory of Dirac equations \cite{Dirac} that a gauge field minimally coupled to a Dirac particle has not only an orbital coupling, that gives rise to a phase difference for different paths, but in addition automatically generates a `Zeeman' coupling of the `spin' of the Dirac particle to the curl of the gauge field. Here, we shall refer to this coupling as the {\it psuedo-Zeeman} coupling since of course the physical spin of the quasiparticle is not involved. Rather, the two component structure of the quasiparticle wavefunction arises from electron-hole mixing. 

Thus, in the linearized theory, in addition to generating the requisite statistical interaction between the vortices and quasiparticles, we have automatically produced a psuedo-Zeeman interaction of the quasiparticle with the curl of the ficticious vector potential  $\vec{a}$.   

Thus, the linearized Hamiltonian (\ref{hamlt}) is not a faithful representation of our original problem. In addition to the requisite statistical interaction, the Franz-Tesanovic transformation automatically hands us a spurious psuedo-Zeeman interaction of the quasiparticles with the curl of the vector potential $\vec{a}$, which has the form of delta functions of strength $q_i\pi$ located at the vortex centers.  As a conseqence, this continuum theory (defined without an appropriate short distance regularization) is not invatiant under different Franz-Tesanovic transformations, i.e. different choices of the $q_i$. This feature has in fact been observed in the numerical studies of \cite{Vafek}. The origin of this problem can be traced back to the singular gauge transformation required to turn the multiple valued wavefunction with branch cuts, to a single valued wavefunction. The transformation is singular at the vortex centers, and since the quasiparticle amplitude (in the continuum theory) is finite at these points the essential physics of the problem is altered by this singular mapping. In addition, due to these psuedo-Zeeman couplings, the linearized continuum theory breaks  ${\mathcal T}_{Dirac}$ in an essential way, and would give rise to spurious gaps in the band structure of the quasiparticles. We believe this to be at the root of some numerical results such as the gaps reported in \cite{Marinelli} for inversion symmetric non-Bravais lattices. This issue is discussed in more detail in the next Appendix.

Clearly, a proper regularization of the problem that takes into account the vortex cores, will avoid this problem. The simplest regularization proceedure would be to imagine that the quasiparticles live on a microscopic lattice with the vortices at the plaquette centers - then the quasiparticles are oblivious to the existence of the flux ($\nabla \times \vec{a}$) of the ficticious gauge field, except for the Aharanov-Bohm phases that they generate.

At this stage it may appear that the conclusion obtained in Section \ref{sect3} of the existence of Dirac nodes for the $hc/2e$ vortex lattice case would also be suspect, since this was argued witin the continuum theory where the spurious psuedo-Zeeman coupling is expected to be present. However, a closer look at the argument reveals that we have implicitly turned off this spurious coupling. This is assumed when making the gauge transformation that combined with ${\mathcal T}_{Dirac}$ (\ref{combination}) leaves the Hamiltonian invariant and protects the Dirac cones. This gauge transformation is only allowed if the quasiparticles are oblivious of the psuedo-Zeeman coupling. Therefore, the results obtained with this argument actually applies to a properly regularized theory (such as with a microscopic lattice) and reflect the physics of the original problem. 

	Since the existence of the Dirac cones in the linearized theory of the $hc/2e$ vortex lattice is one of the central results of this paper, we provide below an alternative argument that leads to this conclusion, but avoids the use of the Franz-Tesanovic transformation.  The linearized Hamiltonian for a vortex lattice of $hc/2e$ vortices before making the F-T transformation:
\begin{eqnarray}
{\mathcal H}^{1}_L \psi' &=& E\psi' \nonumber \\
{\mathcal H}^{1}_L &=& v_F\{p_x\sigmaz + \alpher p_y\sigmax +P_{sx}(\vec{r}) \}
\end{eqnarray}
where the wavefunctions $\psi'$ are not single valued, but have a set of branch cuts originating from the vortex centers, across which the wavefunction changes sign. Now, it is clear that ${\mathcal T}_{Dirac}$ is a symmetry of this problem, since if $\psi'$ is an eigenstate of the Hamiltonian with energy $E$, then $\TAU\psi'^*$ is also an eigenstate with the same energy $E$ and has exactly the same set of branch cuts as $\psi'$. Notice that this symmetry is only valid because the discontinuity across a branch cut corresponds to a sign change, and not a general phase factor that would be altered by complex conjugation. Further, if  we select the branch cuts between vortices in a periodic fashion, we obtain a unit cell with an even number of vortices. Since the superflow is also a periodic function with the same periodicity as the vortex lattice, the eigenstates of ${\mathcal H}^{1}_L$ can be labelled by the a crystal momentum ($\vec{k}$) that takes values within a Brillouin zone that is determined not only by the vortex lattice but also by the choice of branch cuts. Once again, due to  ${\mathcal T}_{Dirac}$ there is a degeneracy of states with crystal momentum $\vec{k}$ and  $-\vec{k}$ and for the special points in the Brillouin zone for which these two are the same, i.e. $\vec{k}\equiv -\vec{k}$(mod $\vec{G}$), there is a pair of degenerate states. This leads to Dirac cones centered at these points in the Brillouin zone. For inversion symmetric lattices, it is always possible to select a set of branch cuts that are taken to themselves on inversion. Consider, for convenience, working with such a set of branch cuts - now it is easily seen that if $\psi'(\vec{r})$ is an eigenstate of the Hamiltonian ${\mathcal H}^{1}_L$ with energy $E$, then $\psi'(-\vec{r})$ is also an eigenstate, with precisely the same set of branch cuts as $\psi'(\vec{r})$, but with energy $-E$. Thus, for inversion symmetric lattices the spectrum at each node is particle-hole symmetric. 

\begin{figure}
\epsfxsize=3.2in
\centerline{\epsffile{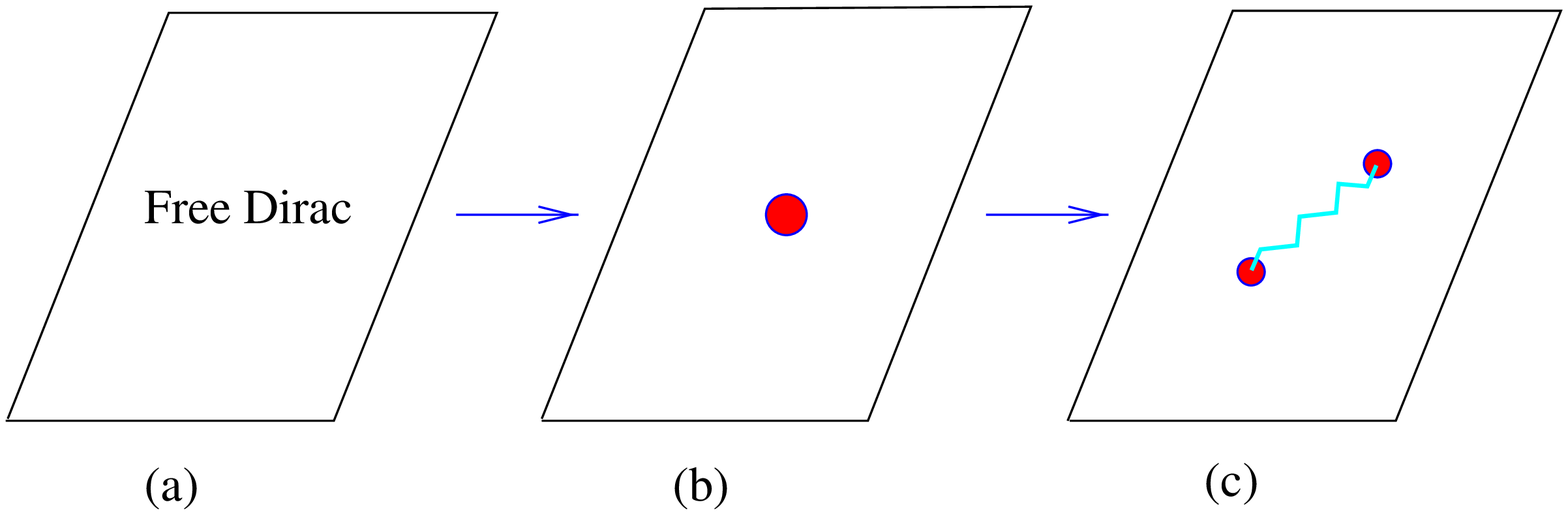}}
\vspace{0.1in}
\caption{Alternate argument for a degenerate pair of states at zero energy, for inversion symmetric vortex lattices of $hc/2e$ vortices. Starting with (a) a free Dirac particle on the torus for which a pair of zero energy states exists, it can be argued from the presence of \Tdirac  and particle-hole symmetry that (b) this pair persists at zero energy when a $hc/e$ vortex is introduced, and (c) continues to exist at zero energy when this double vortex is pulled apart into a pair of $hc/2e$ vortices, with a branch cut running between them.}
\label{cut}
\end{figure}

	We can now argue for the existence of a zero energy doublet at the $\vec{k}=0$ ($\Gamma$) point for this case of inversion symmetric lattices. Let us consider for definiteness the case of an oblique lattice with two vortices per unit cell, with the branch cut chosen to go between them as shown in Figure (\ref{cut}c). Starting with a free Dirac particle on the torus (i.e. the unit cell with periodic  ($\vec{k}=0$) boundary conditions) we note that the spectrum contains a pair of  zero energy states corresponding to the spatially constant spinors. Now, imagine introducing a $hc/e$ (double) vortex at the center of the unit cell as shown in Figure (\ref{cut}b) - as argued previously from the ${\mathcal T}_{Dirac}$ and particle-hole symmetry for this case, a pair of states have to still exist at zero energy. We can now imagine this double vortex being drawn apart into two $hc/2e$ vortices with a branch cut running between then as shown in Figure  (\ref{cut}c). Since we have shown that  ${\mathcal T}_{Dirac}$ and particle-hole symmetries continue to exist for this situation, we conclude once more that a pair of zero energy states exist for a vortex lattice of $hc/2e$ vortices, so long as it posesses inversion symmetry. This pair of states at zero energy is also required to be present to keep the count of the total number of energy states the same as in the free Dirac case, where there were an odd number of doublets.       

	In this appendix we have explored some subtle aspects involved with using the Franz-Tesanovic transformation in the linearized theory. A distinction needs to be drawn between a properly regularized theory, that captures the physics of the original problem and which is the focus of our work, and the naive continuum version of the theory that includes the psuedo-Zeeman coupling and is hence not a faithful rendering of the original problem. As a consequence of the psuedo-Zeeman coupling, this continuum theory is not invariant under different choices of F-T transformations and breaks ${\mathcal T}_{Dirac}$ in an essential way and so does not in general posess Dirac cones in its spectrum.


\section*{Appendix C: Comparison with Numerical Work}
In this appendix we compare the results we have obtained from symmetry considerations of the linearized theory, with the detailed numerical results reported in \cite{Marinelli}\cite{Vafek}. From our analysis we expect to find Dirac cones at the special points in the Brillouin zone for which $\vec{k}$ and $-\vec{k}$
are equivalent, and further, for inversion symmetric lattices we expect a Dirac cone centered at zero energy. As we shall see these expectations are not quite borne out by the numerics. We suspect this to be due to the fact that some of the numerical work is inadvertantly studying the naive continuum theory with spurious psuedo-Zeeman couplings. In order to substantiate this claim we consider in more detail the naive continuum theory, despite the fact that it is not a faithful representation of the original physical problem. As we will see, a detailed correspondence between some of the numerical results and the spectrum of the naive continuum theory can be made. 

As has been noted, the naive continuum limit of the F-T transformed linearized theory involves a spurious psuedo-Zeeman coupling that breaks ${\mathcal T}_{Dirac}$ in an essential way and hence, we do not expect to find Dirac cones (band touchings) in this theory. However, in the presence of other symmetries, some Dirac cones may survive. 
\begin{figure}
\epsfxsize=3.2in
\centerline{\epsffile{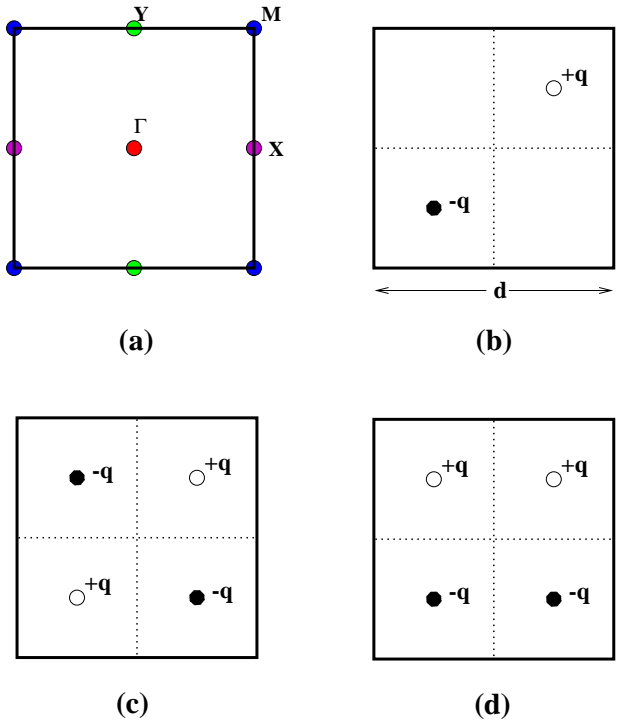}}
\vspace{0.1in}
\caption{(a) The Brillouin Zone for a square vortex lattice. (b) A unit cell with two vortices as considered in reference [11]. (c) and (d) unit cells as considered in reference [12]}
\label{fig5}
\end{figure}
Consider for example a square lattice of $hc/2e$ vortices within a F-T gauge choice that leads to a unit cell with with two vortices as shown in Figure (\ref{fig5}b) (and as considered in \cite{Marinelli}). The Hamiltonian for this situation is:

$$
\label{haml1}
H^{FT}_1 = v_F\{(p_x + a_x)\sigmaz + \alpher (p_y+a_y)\sigmax+\hat{x}\cdot\vec{P}_s(\vec{r})\}
$$
with the solenoid fluxes:
\begin{eqnarray*}
\vec{\nabla}\times \vec{a}(\vec{r}) &=& \hbar\pi \sum_{m,n\in {\mathcal Z}} q [\delta(\vec{r}-d(m+\frac14)\hat{x}-d(n+\frac14)\hat{y})\\
	& & -\delta(\vec{r}-d(m-\frac14)\hat{x}-d(n-\frac14)\hat{y})]
\end{eqnarray*}

Now, while ${\mathcal T}_{Dirac}$ is not a symmetry of this problem, due to the doubling of the unit cell, there is a generalized `glide' symmetry - a combination of ${\mathcal T}_{Dirac}$ and a translation by half a special lattice vector - that {\it is} present. Thus, consider the transformation,
\be
\label{glide}
\psi(\vec{r}) = \TAU \psi^*(\vec{r}+\frac{d}2\hat{x}+\frac{d}2\hat{y})
\ee

. The Dirac time reversal operation reverses the sign of the vector-potential $\vec{a}$, but the translation by half a diagonal lattice vector interchanges the $q$ and $-q$ fluxes and hence restores invariance. Of course, the superflow is unaffected by this translation as it is periodic with a smaller unit cell.  This leads to a degeneracy between energy levels with crystal momentum $\vec{k}$ and $-\vec{k}$ and for those points in the Brillouin zone that are taken to themselves under this operation (labelled $\Gamma$, X, Y and M in Figure \ref{fig5}a). At these points we have a pair of wavefunctions $\psi_{\vec{k}}(\vec{r})$ and $\TAU\psi^*_{\vec{k}}(\vec{r}+\frac{d}2(\hat{x}+\hat{y}))$ with the same energy eigenvalue. In order that we have a degenerate doublet, we need that this pair be linearly independent. Earlier, when the symmetry operation was simply ${\mathcal T}_{Dirac}$, the linear independence followed from the fact that the two wavefunctions were othogonal from the antisymmetry of $\TAU$. Here however we will need to adopt a more indirect line of reasoning, paralleling that used to establish Kramer's degeneracy for electrons \cite{Sakurai}. To learn whether these wavefunctions are indeed linearly independent at these special points in the Brillouin zone (for $\vec{k}\in \{0,\frac{2\pi}d\hat{x},\frac{2\pi}d\hat{y},\frac{2\pi}d(\hat{x}+\hat{y})\}$ the $\Gamma$, X, Y and M points), we adopt a proof by contradiction. Assume that the pair of functions {\it are} linearly dependent. Then, since they have the same normalization, they can at best differ by a phase factor $\mbox{e}^{i\chi}$.
\be
\label{dependence}
\psi_{\vec{k}}(\vec{r})=\mbox{e}^{i\chi} \TAU \psi^*_{\vec{k}}(\vec{r}+\frac{d}2(\hat{x}+\hat{y}))
\ee 

Performing the `glide' symmetry operation on both sides of the equation we obtain:
\be
\label{lhs}
\TAU\psi^*_{\vec{k}}(\vec{r}+\frac{d}2(\hat{x}+\hat{y})) = -\mbox{e}^{-i\chi}\psi_{\vec{k}}(\vec{r}+d(\hat{x}+\hat{y}))
\ee
where we have used the fact that $\TAU^2 = -\One$. Using now the fact that translation by $d(\hat{x}+\hat{y})$ amounts to a lattice translation within this expanded two vortex unit cell 
$$
\psi_{\vec{k}}(\vec{r}+d(\hat{x}+\hat{y})) = \mbox{e}^{i\vec{k}\cdot(\hat{x}d+\hat{y}d)}\psi_{\vec{k}(\vec{r})}
$$
and substituting equation (\ref{dependence}) for the left hand side of (\ref{lhs}) we obtain:
\be
\mbox{e}^{-i\chi}\psi_{\vec{k}}(\vec{r})=-\mbox{e}^{-i\chi}\mbox{e}^{i\vec{k}\cdot(\hat{x}d+\hat{y}d)}\psi_{\vec{k}}(\vec{r})
\ee
which for the X and Y points ($\vec{k}\in\{\frac\pi{d}\hat{x},\frac\pi{d}\hat{y}\}$) yields the consistent equation $\psi_{\vec{k}}(\vec{r})=\psi_{\vec{k}}(\vec{r})$ whereas for the $\Gamma$ and M points ($\vec{k}\in\{0,\frac\pi{d}\hat{x}+\frac\pi{d}\hat{y}\}$) this yields a contradicition $\psi_{\vec{k}}(\vec{r})=-\psi_{\vec{k}}(\vec{r})$. Thus we can conclude that the pair of states at the $\Gamma$ and M points have to be linearly {\it independent} and the spectrum at these points in the Brillouin zone consists of degenerate doublets. For the X and Y points, this conclusion cannot be drawn. As a consequence, for the situation in Figure \ref{fig5}b, we expect Dirac cones at the $\Gamma$ and M points but not necessarily at the X and Y points, in the continuum theory band structure. Further, inversion symmetry of this vortex lattice gives us a particle hole symmetric band structure \cite{Marinelli} - if $\psi(\vec{r})$ is an eigenstate with energy $E$, then $\TAU\psi^*(-\vec{r})$ is an eigenstate with energy $-E$. As we have reasoned several times before, the particle hole symmetry combined with the doubly degenerate nature of the spectrum at the $\Gamma$ point leads to a Dirac cone centered at zero energy. This is exactly the result of the numerical studies in \cite{Marinelli} for this case. Dirac cones are found at the $\Gamma$ and M points, but not at the X and Y points of the Brillouin zone and at the $\Gamma$ point there is a Dirac cone centered at zero energy. 

A similar analysis for the situation of Figure (\ref{fig5}c and d) can be carried out, and the continuum theory in these two cases are expected to have Dirac cones at the $\Gamma$, X and M points and $\Gamma$, X and Y points respectively as well as a Dirac cone centered at zero energy in both cases. This is exactly as seen in the numerical work on the linearized model in \cite{Vafek}.
\begin{figure}
\epsfxsize=3.2in
\centerline{\epsffile{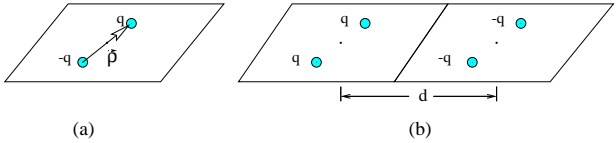}}
\vspace{0.1in}
\caption{An oblique vortex lattice of $hc/2e$ vortices with two vortices per unit cell. In (a) a F-T transformation that preserves this periodicity while in (b) a F-T transformation that doubles the unit cell. Case (b) can be shown to have a pair of zero energy states even in the continuum (unregularized) theory, that are protected by a `glide' symmetry resulting from the artificial doubling of the unit cell.}
\label{fig4}
\end{figure}

If instead we considered a case where such a `glide' symmetry was absent, then in the continuum theory we would not obtain Dirac cones anywhere in the Brillouin zone. For example if we have a vortex lattice that contains two vortices per unit cell, and make an F-T transformation that preserves this periodicity, as for example in Figure \ref{fig4}a we do not expect to find Dirac cones and there will be no states at zero energy. This is also the result obtained in \cite{Marinelli} for the case of a non-Bravais lattice with two vortices per unit cell, within a F-T transformation that preserves this periodicity. If however we were to consider this same problem but within a F-T transformation that artificially doubles the unit cell, as shown for example in Figure \ref{fig4}c, then a `glide' symmetry will be obtained. By the arguments in this Appendix we expect to find Dirac nodes and a pair of zero energy states in this continuum theory. The existence of the zero energy states in this case was explicitly verified in Appendix A within perturbation theory, where the `glide' symmetry protects the zero energy states for all values of the flux $q$ (not just odd integers) and allows it to be accessed within perturbation theory. This is a clear indication of the non-invariance of the continuum theory to different Franz-Tesanovic gauge transformations.

Such non-invariance of the spectrum to different FT transformations was reported in \cite{Vafek} where a reciprocal space approach was used to numerically diagonalize the linearized Hamiltonian. Clearly, in this scheme it is the continuum theory which is being studied. Thus, the unphysical psuedo-Zeeman coupling will play a role and further, due to the cutoff in momentum space the solenoids of ficticious flux will acquire a finite thickness and have an effect beyond just implementing the Aharanov-Bohm phase factor of (-1). Hence the non-invariance with respect to different F-T transformations and the detailed agreement with our expectations for the continuum theory. On the other hand the tight binding lattice approach also reported in the same work \cite{Vafek}, which does not adopt a linearization was, naturally, found invariant to different choices of F-T transformations. It is less clear, however, as to why the real space diagonalization approach of \cite{Marinelli} also appears to be simulating the continuum theory. We suspect this could arise if the solenoids of flux happenned to be smoothed out over over a few plaquettes - which would have the same qualitative effect as the presence of a psuedo-Zeeman coupling. In principle however, such a real space approach should be able to provide an appropriate regularization for the F-T transformed linearized problem. One particularly convenient way to implement the (-1) phase factors would be to use Ising link variables.
   
In summary, for the F-T transformed linearized theory, it is important to draw a distinction between a properly regularised theory and the naive continuum limit. The latter is plagued by the presence of the psuedo-Zeeman coupling that renders it non-invariant under different F-T transformations, breaks ${\mathcal T}_{Dirac}$ in an essential way and is not a faithful representation of the original problem. We suspect that some of the results reported in the pioneering numerical work on this linearized approximation \cite{Marinelli}\cite{Vafek} may have actually been studying this continuum theory. In order to substantiate this, in this appendix we have analysed the continuum theory using the methods of this paper, and drawn conclusions regarding the nature of the spectrum, that are found to agree well with some of these numerical results. 

\section*{Appendix D: Chern Numbers for Quasiparticle Bands}
Consider calculating the spin Hall conductivity of the full quasiparticle Hamiltonian $H_{BdG}$ (the linearized Hamiltonian has vanishing spin Hall conductance due to invariance under \Tdirac). We use the Kubo formula that evaluates the response of the quasiparticle current $j_x=\frac{\partial H_{BdG}}{\partial p_x}$ to a perturbation corresponding to a uniform Zeeman gradient (e) $\delta H=e y \One$ along the orthogonal direction. 

At zero temperature the spin Hall conductivity for an isolated filled band takes a particularly simple form, just as in the case of the electrical Hall effect with charged particles in a periodic potential \cite{TNNK}. It is proportional to a topological invariant, the Chern number \cite{Morita} that can be asigned to the quasiparticle band labelled by `n' via the formula:
\begin{eqnarray}
\sigma^{spin}_{xy} (n) &=& \frac1h (\frac{\hbar}2)^2 C_n \\
C_n &=& \frac1{2\pi}\int_{B.zone} \vec{\nabla}_k \times \vec{{\mathcal A}}(k) d^2k 
\label{chern}
\end{eqnarray}
The vector potential is defined in terms of the wavefunctions of the quasiparticles in that band by 
\be
\vec{{\mathcal A}}(k) = -i \int \frac{d^2r}{{\mathcal V}} \{ u^*_{nk}(r)\vec{\nabla}_k u_{nk}(r)+v^*_{nk}(r)\vec{\nabla}_k v_{nk}(r)\}
\label{connection}
\ee

where  the integral is over the unit cell, and we have written out the components of the quasiparticle wavefunctions of band `n' as $\psi_{nk}(r)=(u_{nk}(r) \; v_{nk}(r))^{T}$ that we choose to be normalized according to:
\be
1=\int_{unit-cell} \frac{d^2r}{{\mathcal V}}(|u_{nk}(r)|^2 + |v_{nk}(r)|^2)
\ee
where ${\mathcal V}$ is the area of the unit cell. Clearly, a phase redefinition of the wavefunction at each point of the Brillouin zone $\psi_{nk} \rightarrow \mbox{e}^{i\chi(k)}\psi_{nk}$ corresponds to a gauge transformation $\vec{{\mathcal A}}(k) \rightarrow \vec{{\mathcal A}}(k) + \vec{\nabla}_k \chi(k)$. This does not affect the value of the Chern number - which is thus defined despite the phase ambiguity of the wavefunctions. Following standard arguments developed for electrons \cite{Kohmoto}, it is easily established that for the case of superconductor quasiparticles as well the $C_n$ are integers and hence the spin Hall conductivity, in appropriate units, will be quantized to integers for this case of quasiparticles of a spin-singlet superconductors, so long as the chemical potential for quasiparticles lies in a gap \cite{Volovik}\cite{Senthil}. For the case of homogenous superconductors that break Time reversal and Parity symmetries, the appropriate limit of experssion (\ref{chern},\ref{connection}) reduces to the topological invariant of \cite{Volovik}\cite{Read}. 

The quasiparticles therefore, could be said to blend together Bloch wave and Landau level properties.  While they form bands that are labelled by Bloch crystal momenta, these bands, like Landau levels, carry a Chern number that is related to the spin-Hall conductivity.    


\begin{thebibliography}{x}
\bibitem{tmax} At temperature T, the largest typical momentum for $\alpha > 1$ is $[\delta p_y]_{max}=\alpha T/v_F$. The, the quadratic term is also of the same order if the temperature reaches $T_{max}= \frac1{2m}[\delta p_y]^2_{max}=4 E_F/\alpha^2$. For YBa$_{2}$Cu$_{3}$O$_{6.9}$, $\alpha = 14$, $E_F=3,300K$ and also there is a mass anisotropy that we have not explicitly included which boosts $T_{max}$ by a factor of $\sim 3$ to give eventually $T_{max} \sim 200$Kelvin.
\bibitem{CS}
Notice, that each of these factors an also be written as $(\frac{z-z_i}{|z - z_i|})^{\frac12 q_i}$ which is familiar from Chern-Simons transformations for the QHE problem.
\bibitem{ptclhole} In fact, particle-hole symmetry may be obtained at every point on the Brillouin zone \cite{Marinelli}, by combining the previous inversion transformation with ${\mathcal T}_{\mbox{Dirac}}$ . Thus, the transformation 
$\psi(x,y) \rightarrow \TAU \psi^*(-x,-y)$  leaves the crystal momentum unaffected, while changing the sign of the energy.
\bibitem{Odd} Strictly speaking we just need to have an odd number of doublets at zero energy, which in the generic case will be just one doublet.
\bibitem{orthogonal} This is guaranteed at node $1$, $\bar{1}$ ($2$, $\bar{2}$) if the vortex lattice is symmetric under the reflection $x \rightarrow -x$ ($y \rightarrow -y$).
\bibitem{antinodemD} In fact, this equality of $m_D$ for the opposite nodes may be proved more generally, and holds even in the absence of inversion symmetry. It derives from the intrinsic particle hole symmetry present in the Boguliubov-de Gennes equation, that relates the opposite nodes.
\end{thebibliography}
\end{document}